\documentclass[a4paper,11pt]{article}
\usepackage{jheppub} % for details on the use of the package, please see the JINST-author-manual
\usepackage{common}
\usepackage{comment}
\usepackage{lineno}
\usepackage{xspace}
\usepackage{subcaption}
\usepackage[T1]{fontenc}
\usepackage{multirow}
\usepackage{float}
\usepackage{siunitx}

\title{\boldmath Expected Sensitivity of the Light Dark Matter eXperiment to Long-Lived Dark Photons and Axion-Like Particles}

\collaboration{LDMX Collaboration}
\collaborationImg{\includegraphics{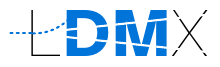}}

\author[a]{Torsten~Akesson}
\author[a]{Clay~Barton}
\author[b]{Charles~Bell}
\author[c]{Elizabeth~Berzin}
\author[d]{Liam~Brennan}
\author[a]{Lene~Kristian~Bryngemark}
\author[e]{Lincoln~Curtis}
\author[f]{Patill~Daghlian}
\author[e]{E.~Craig~Dukes}
\author[g]{Valentina~Dutta}
\author[f]{Bertrand~Echenard}
\author[e]{Ralf~Ehrlich}
\author[h]{Thomas~Eichlersmith}
\author[a]{Einar~El\'{e}n}
\author[h]{Andrew~Furmanski}
\author[c]{Majd~Ghrear}
\author[i]{Matthew~Gignac}
\author[i]{Matt~Graham}
\author[d]{Chiara~Grieco}
\author[e]{Craig~Group}
\author[a]{Hannah~Herde}
\author[b]{Christian~Herwig}
\author[f]{David~G.~Hitlin}
\author[e]{Tyler~Horoho}
\author[d]{Joseph~Incandela}
\author[f]{Nathan~Jay}
\author[j]{Wesley~Ketchum}
\author[j]{Gordan~Krnjaic}
\author[d]{Oscar~Lewis}
\author[d]{Yuze~Li}
\author[h]{Jeremiah~Mans}
\author[e]{Cristina~Mantilla~Suarez}
\author[d]{Sanjit~Masanam}
\author[h]{Steven~Metallo}
\author[f]{Sophie~Middleton\footnote{Corresponding author}}
\author[i]{Timothy~Nelson}
\author[c]{Rory~O'Dwyer}
\author[f]{James~Oyang}
\author[g]{Pritam~Palit}
\author[e]{Jessica~Pascadlo}
\author[i]{Emrys~Peets}
\author[a]{Luis~Sarmiento~Pico}
\author[a]{Ruth~Pöttgen}
\author[h]{Chelsea~Rodriguez}
\author[d]{Lincoln Satterthwaite}
\author[i]{Philip~Schuster}
\author[e]{Matt~Solt}
\author[c]{Lauren~Tompkins}
\author[i]{Natalia~Toro}
\author[j]{Nhan~Tran}
\author[d]{Tamas~Vami}
\author[e]{Kieran~Wall}
\author[a]{Erik~Wallin}
\author[j]{Andrew~Whitbeck}
\author[d]{Jihoon~Yoo}
\author[d]{Danyi~Zhang}

\affiliation[a]{Lund University, Department of Physics, Box 118, 221 00 Lund, Sweden}
\affiliation[b]{University of Michigan, Ann Arbor, MI 48109, USA}
\affiliation[c]{Stanford University, Stanford, CA 94305, USA}
\affiliation[d]{University of California, Santa Barbara, Santa Barbara, CA 93106, USA}
\affiliation[e]{University of Virginia, Charlottesville, VA 22904, USA}
\affiliation[f]{California Institute of Technology, Pasadena, CA 91125, USA}

\affiliation[g]{Carnegie Mellon University, Pittsburgh, PA 15213, USA}
\affiliation[h]{University of Minnesota, Minneapolis, MN 55455, USA}
\affiliation[i]{SLAC National Accelerator Laboratory, Menlo Park, CA 94025, USA}

\affiliation[j]{Fermi National Accelerator Laboratory, Batavia, IL 60510, USA}

\emailAdd{smidd@caltech.edu}

\abstract{
The Light Dark Matter eXperiment (LDMX) is an electron-beam fixed-target experiment primarily designed to achieve world-leading, model-independent sensitivity to sub-GeV dark matter particles.
LDMX aims to identify dark sector particle production through the detection of events with substantial missing energy and momentum, a signature of invisible particles escaping detection.
Beyond this primary objective, LDMX offers a complementary search strategy for long-lived, visibly decaying particles, such as dark photons and axion-like particles.
We present the first detailed evaluation of the ability of LDMX to identify visibly decaying, long-lived particles that couple to electrons using a detailed simulation, based on the {\sc Geant4}-toolkit, that incorporates realistic detection efficiencies and background levels.
We demonstrate that LDMX can achieve a sensitivity that is competitive with other experiments that are currently running.
The models explored in this paper are distinct and complementary to those probed in the LDMX flagship missing-momentum analysis.
Through searching for both invisible dark matter and visibly decaying long-lived signatures, LDMX will significantly advance the search for light dark matter and provide a broad exploration of the sub-GeV dark sector.}

\begin{document}
\maketitle
\flushbottom
\clearpage

\section{Introduction and Motivation}\label{sec:intro}

\begin{figure}[h!]
	\centering
	\includegraphics[width=0.99\linewidth]{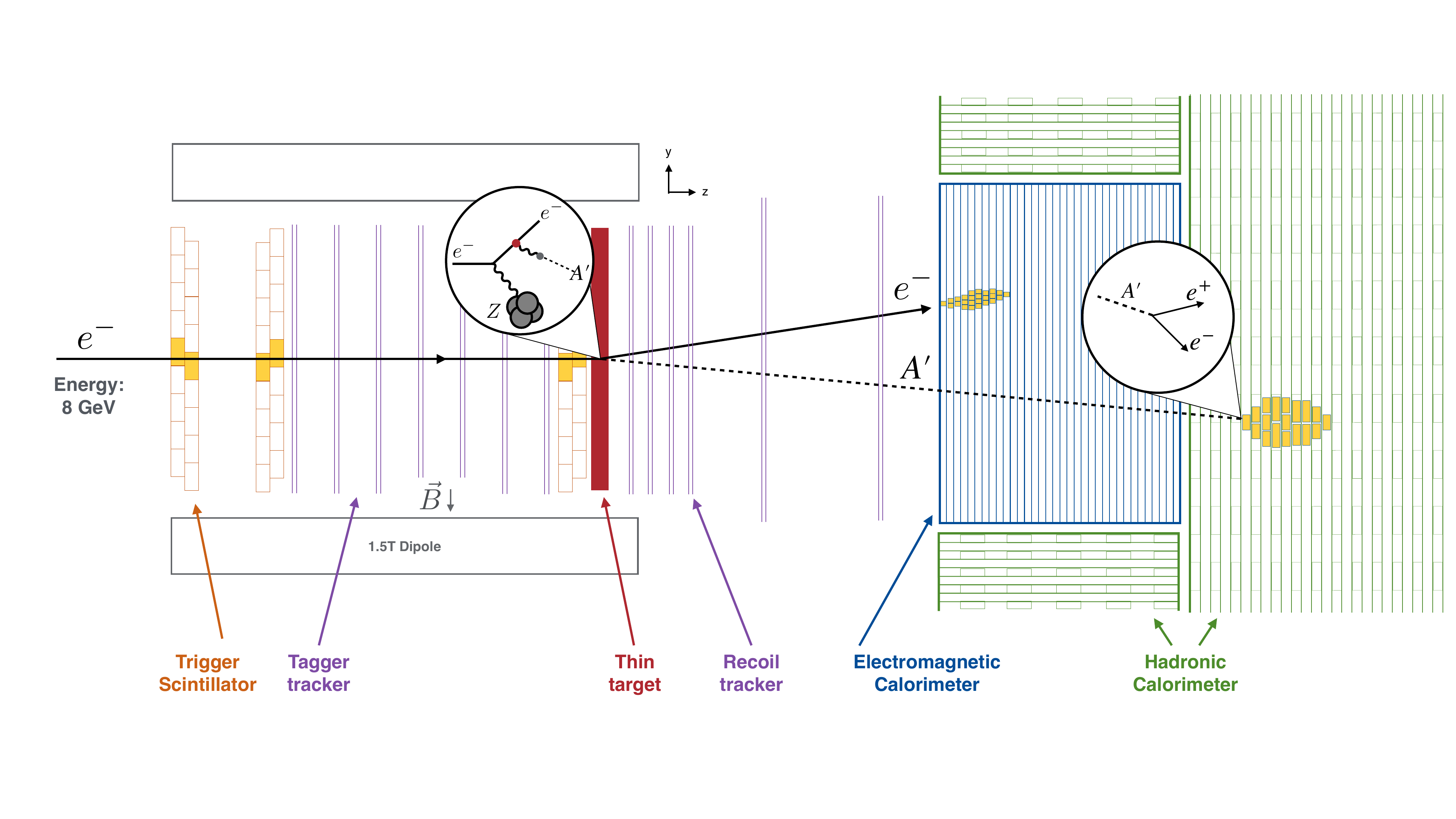}
	\caption{A plan view of the LDMX apparatus with visibly decaying $A'$ producing a displaced decay vertex within the HCal overlaid and subsystems labeled. An analogous sketch can be shown for the visibly decaying ALP.}
	\label{fig:ldmx_visibles_cartoon}
\end{figure}

The Light Dark Matter eXperiment (LDMX) is primarily designed to search for dark matter (DM) production via a missing energy and momentum signature~\cite{TDR, LDMX:2018cma}. To effectively identify this signal, all Standard Model backgrounds that could mimic a missing-momentum signature must be vetoed with very high efficiency \cite{LDMX:2023zbn}.
This requirement is the driving force behind the design of the LDMX apparatus, a schematic diagram of which is presented in Figure~\ref{fig:ldmx_visibles_cartoon}.
The design includes a silicon tracking system upstream and downstream of a thin tungsten target, an electron counting trigger scintillator system, an electromagnetic calorimeter (\ecal), and a hadronic calorimeter (\hcal).
A comprehensive overview of the LDMX experiment can be found in \cite{TDR}.
LDMX will take its data with an 8 GeV electron beam extracted from the Linac Coherent Light Source II (LCLS-II) at SLAC \cite{Raubenheimer:2018wwc} using the Linac to End Station A (LESA) beamline~\cite{LESA}.
The total LDMX dataset is anticipated to be $10^{16}$ electrons on target (EoT), and an initial pilot run will take $4 \times 10^{14}$ EoT.

As a result of its unique design, LDMX is capable of probing physics signatures beyond the missing-momentum dark matter analysis. LDMX functions as a fully instrumented beam dump with large acceptance and, thus, is well-suited for searching for visible decays of exotic particles.
LDMX's sensitivity will be optimal for particles which are produced in the target and that have lifetimes long enough to reach the hadronic calorimeter $\sim$90 -- 580\,cm from the target.
Preliminary sensitivity estimates for models of visibly decaying dark photons ($A'$) and axion-like particles (ALP or $a$) have been derived based on theoretical calculations and an early design of the LDMX detector~\cite{Berlin:2018bsc}.
In this paper, we present a comprehensive simulation study that evaluates the ability of LDMX to identify long-lived particles (LLPs) that decay to an $e^+e^-$ pair in the \hcal.
We consider two explicit, minimal, well-motivated, LLP models (the $A'$ and ALP) to showcase LDMX's potential reach for such a signature.

\begin{figure}
    \centering
    \includegraphics[width=0.9\linewidth]{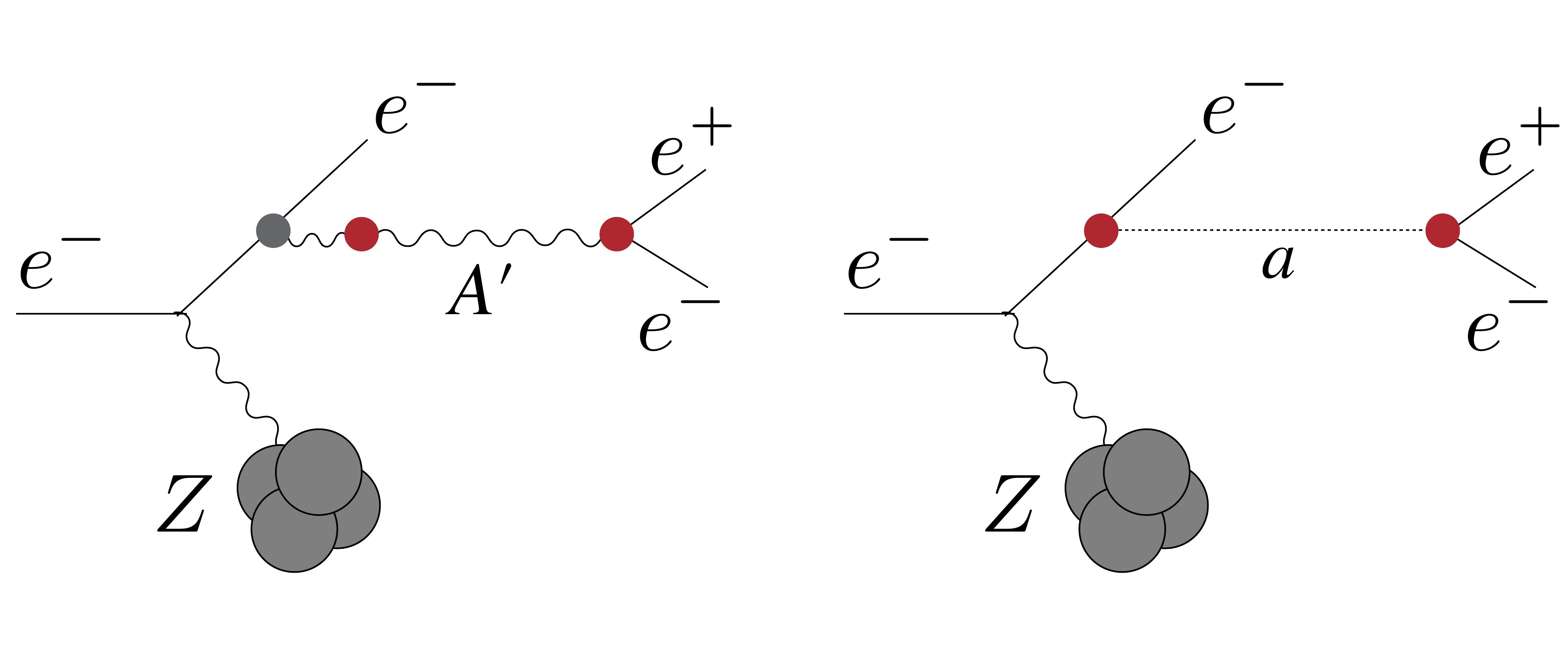}
    \caption{Feynman diagrams for the production and decay of an $A'$ (left) and an ALP ($a$) (right).}
    \label{fig:feynmandiagram}
\end{figure}

Feynman diagrams for the production and decay of both LLP models considered are shown in Figure~\ref{fig:feynmandiagram}.
For the minimal $A'$ model, the $A'$ interacts with the Standard Model through kinetic mixing~\cite{holdom1985} with the photon.
The kinetic mixing parameter between the $A'$ and the Standard Model photon can be denoted as $\varepsilon$.
If $2m_e < m_{A'} < 2m_\chi$, where $m_\chi$ is the mass of the dark matter particle, the $A'$ will primarily decay to an electron-positron pair.
ALPs are predicted by many beyond Standard Model theories and are well-motivated dark sector candidates.
ALPs could also be produced through interactions between the beam and the LDMX target and could also decay visibly within the LDMX detectors, with displaced vertices producing particles in the HCal.
ALPs could be produced in LDMX through a coupling to photons or electrons, with $g_{a \gamma}$ and $g_{a e}$ denoting the ALP- photon and electron couplings, respectively.
Production of ALPs through photon couplings generally results in more energetic recoil electrons, which requires a different triggering scheme from the one described below.
Therefore, this paper will concentrate on the case where the electron coupling is dominant, projected sensitivity to ALP-$\gamma$ with an upgraded trigger scheme is presented in \cite{TDR}.

Although the physics governing these two signal modes ($A'$ and ALP) differ, the signatures within the LDMX detectors will be similar.
Consequently, in this article, we present a single analysis procedure and apply it to both signal searches.
We present a reach estimate for a visibly decaying, displaced, $A'$ or ALP using a full detector simulation for the expected run schedule of an \qty{8}{\GeV} electron beam with \num{e14} EoT.
The results of \num{e14} EoT are extrapolated to estimate the sensitivity for the full anticipated dataset of \num{e16} EoT.

\section{Detector Components}

\begin{figure}
    \centering
    \includegraphics[width=0.44\linewidth]{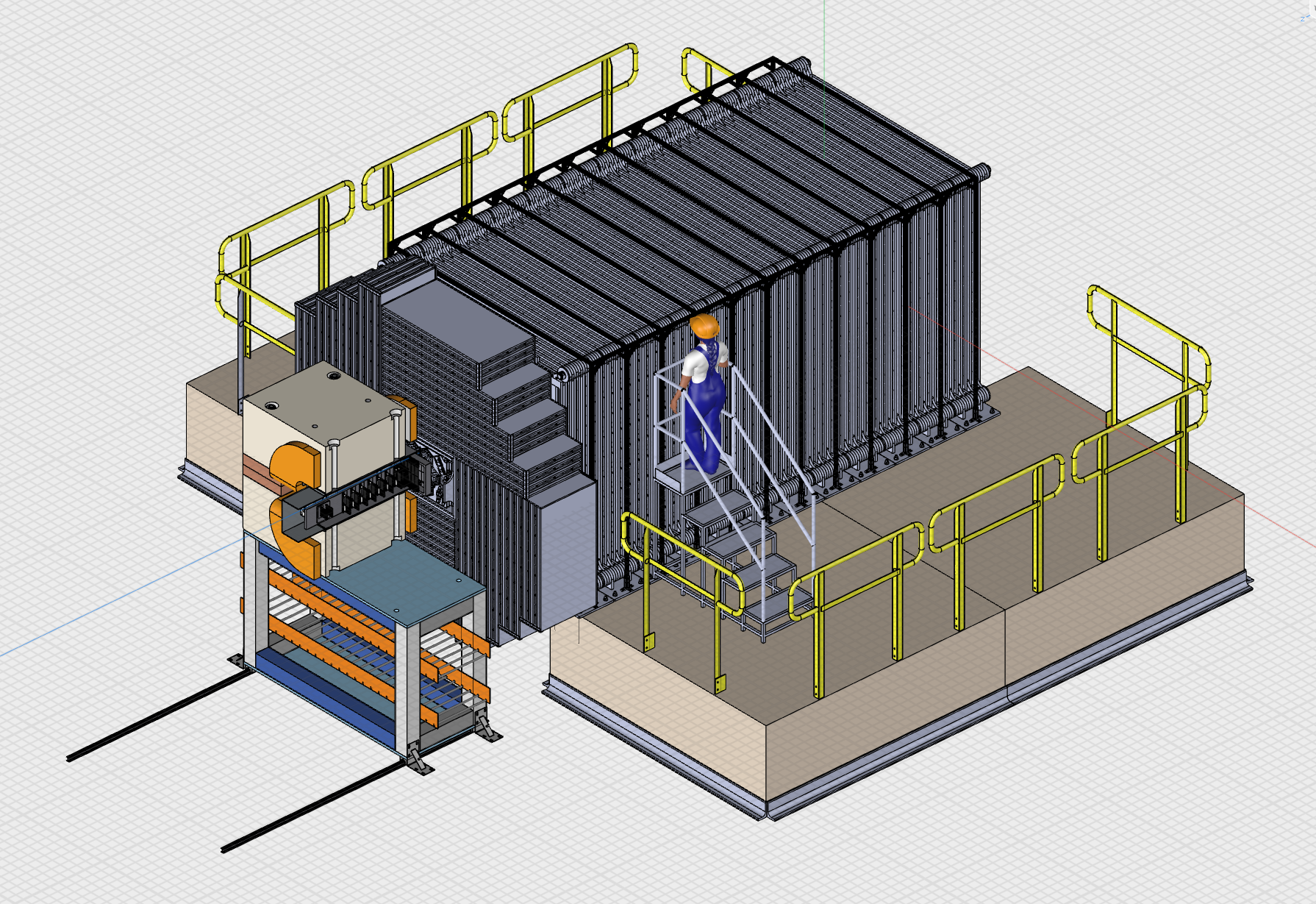}
    \includegraphics[width=0.54\linewidth]{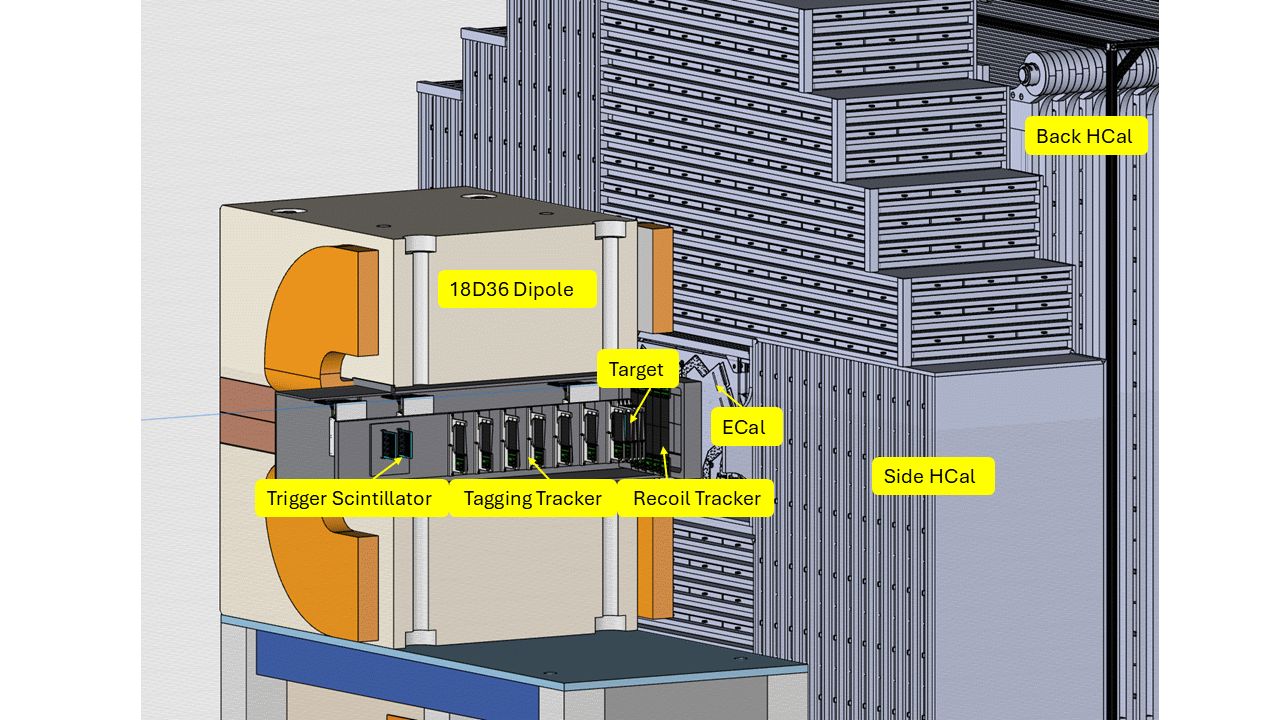}
    \caption{A model of the complete LDMX detector apparatus with a human for scale (left) and a partial cutaway to show various sub-detectors (right).}
    \label{fig:detector}
\end{figure}

A mechanical mock-up of the full detector apparatus is shown in Figure~\ref{fig:detector}.
LDMX will utilize a low current, high duty-factor 8 GeV electron beam fired at a thin (0.1 $X_{0}$) tungsten target.
The tracking system consists of one silicon-strip tracker upstream of the target (the tagging tracker) and one downstream of the target (the recoil tracker) to measure the multiplicity, positions, and four-momenta of charged particles with $p \geq 50$\,MeV.
Downstream of the tracker is the \ecal, which measures the energy of the recoil electron and other interaction products.
The \ecal is a high-granularity, high-density silicon-tungsten sampling calorimeter, capable of full electromagnetic shower containment at 8 GeV and resolving individual products from the target interaction at \num{e16} EoT .
The \ecal is based on the design of the CMS High Granularity Calorimeter for the CMS Phase-2 upgrade~\cite{CMS-HGCal-2017}.

The \hcal is a steel-scintillator sampling calorimeter using the scintillator counter design from Mu2e's cosmic ray veto~\cite{mu2e2014}.
Each scintillator counter is 50\,mm wide and 20\,mm thick and is co-extruded with a TiO$_2$ reflector coating.
A 1.8\,mm diameter wavelength-shifting fiber is threaded through each counter.
Light produced in the scintillator is collected by the wavelength-shifting fiber and read out via silicon photomultipliers on both ends of the counter.
Each 2\,m-by-2\,m \hcal layer consists of 40 doped polystyrene scintillator counters glued to a 25-mm thick steel absorber.
The counter orientation alternates between layers.
The portion of the \hcal surrounding the top, bottom, and sides of the \ecal is called the Side \hcal, while the portion behind the \ecal is called the Back \hcal.
LLPs are typically emitted forward in the detector, so we only consider the Back \hcal in this analysis.
The Back \hcal begins at 87\,cm from the target and extends to 587\,cm.

\section{Simulated Event Samples}\label{sec:simulation}

LDMX uses a purpose-built C++-based software framework, called \ldmxsw, for event processing and simulation.
A software bus model is implemented for communication between data processing modules in sequence.
The data processing pipeline is configured using an embedded Python interpreter at run-time, which allows for the dynamic loading of simulation, digitization, reconstruction, and analysis modules.
The propagation of particles through the detector and their interactions with material are described using a custom version of {\sc Geant4}~\cite{geant4_2003,geant4_2006,geant4_2016} 10.02.p03 incorporating patches to nuclear interaction cross sections and kinematics tailored to the beam energy at LDMX~\cite{LDMX:2018cma}.
Sub-detector specific algorithms are used to digitize and reconstruct data.
Further details on the software framework for LDMX can be found in~\cite{TDR}.

\subsection{Visibly Decaying \texorpdfstring{$A'$}{A'} and ALP Simulations}\label{subsec:signal_simulation}

The kinematics for the production and decay of LLPs studied in this work were simulated using {\sc MadGraph/MadEvent} v3.6.2 ~\cite{alwall2014,ambrogi2018}.
This produces the momenta of the recoil electron and LLP, the decay position of the LLP, and the momenta of the $e^+e^-$ pair from the LLP decay.
Kinematics for four $A'$ masses (5, 10, 50, and 100\,MeV) and three ALP masses (10, 50, and 100\,MeV) were simulated.
These masses were chosen based on initial acceptance studies and theory estimates on LDMX sensitivity to these models~\cite{Berlin:2018bsc}.
The particle kinematics of the {\sc MadGraph/MadEvent} simulation were then passed into \ldmxsw, where {\sc Geant4} simulates the detector response to the recoil electron and the $e^+e^-$ pair, and the data reconstruction is performed.

\subsection{Background Simulation}\label{subsec:background_simulation}

The observable signature for a visibly decaying $A'$ or ALP with a displaced vertex is characterized by an energy deposition near the beam energy in the \hcal.
There are several background processes that can mimic this signature and require effective mitigation.
Backgrounds of concern for this analysis produce a bremsstrahlung photon in the target, which then undergoes another reaction to leave a large energy deposition in the \hcal.
Four background samples were generated for this analysis: photo-nuclear (PN) reactions in the target and in the \ecal, and muon photoproduction in the target and in the \ecal.
Sample sizes equivalent to $1\times10^{14}$ EoT were generated for each of these samples except target muon photoproduction, for which $1\times10^{15}$ EoT equivalent was simulated.
Due to computational constraints, larger samples could not be generated.
Further details on how these samples were generated can be found in~\cite{TDR}.

\section{Trigger and Event Selection}\label{sec:selection}

\begin{figure}
    \centering
    \includegraphics[width=0.48\linewidth]{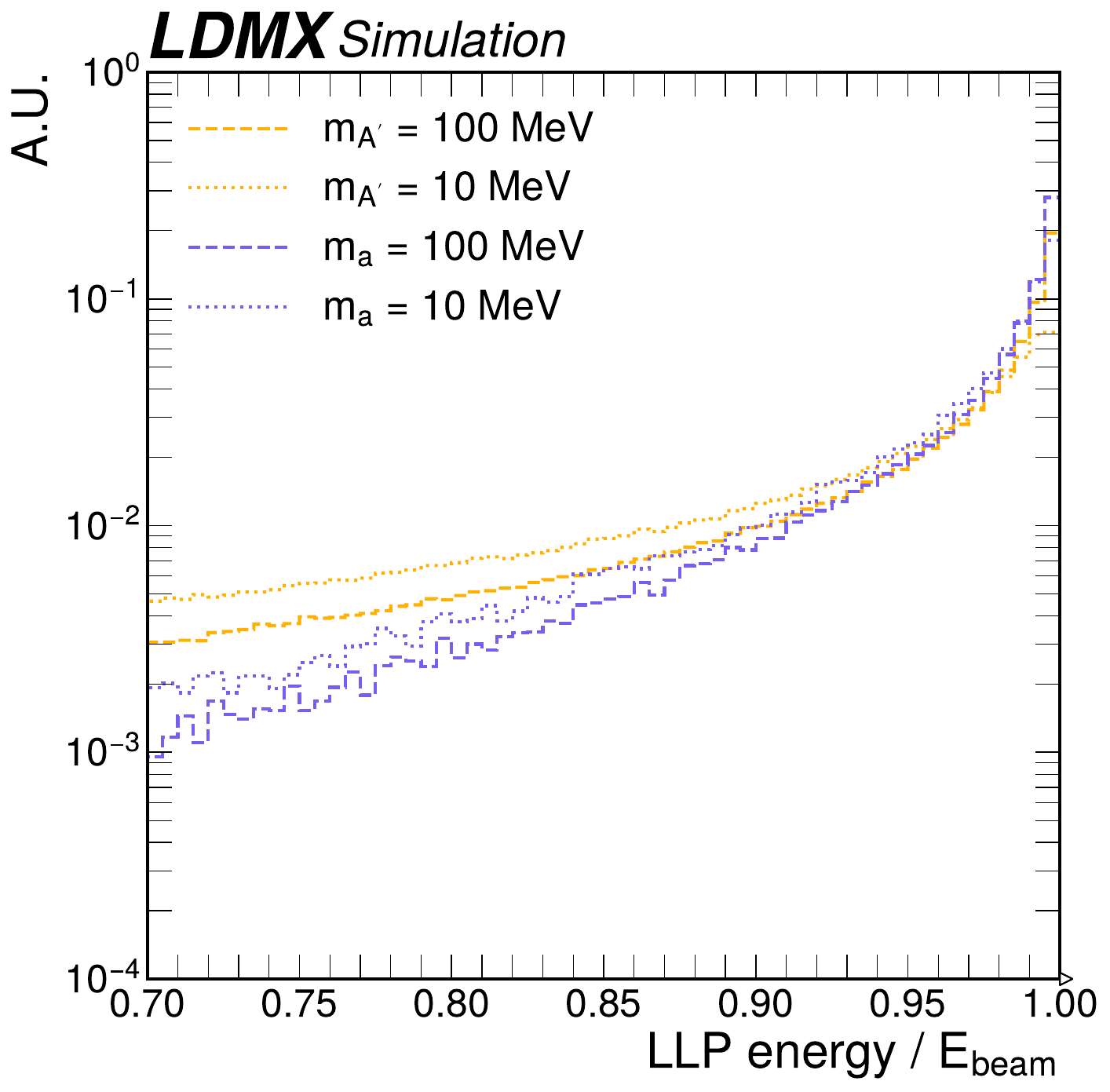}
        \includegraphics[width=0.48\linewidth]{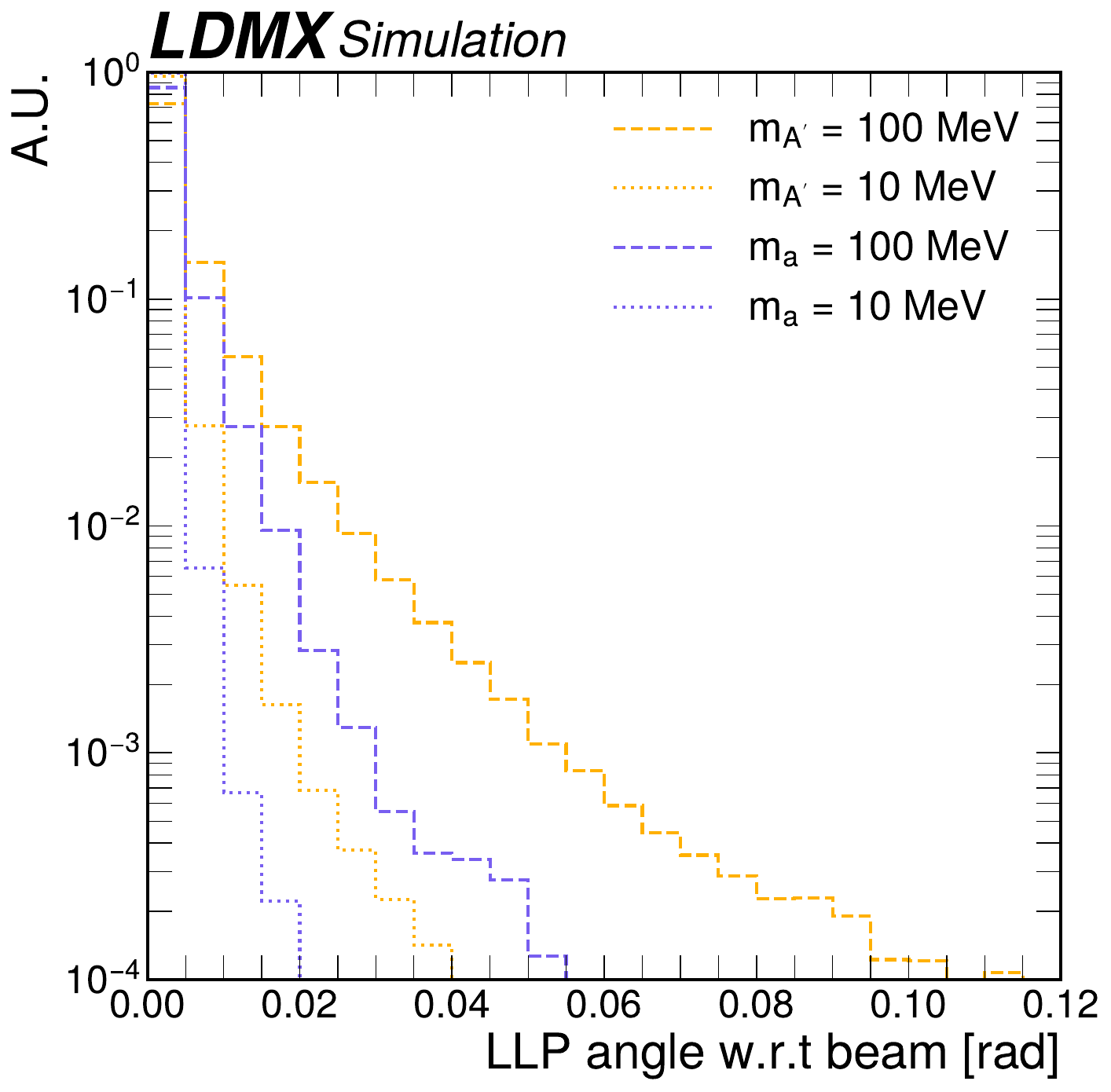}
    \caption{The fraction of beam energy carried away by the LLP (left). The angle of the LLP momentum with respect to the beam axis (right). The kinematic differences are a result of differences in the production of a dark photon or ALP and are explained in detail in \cite{PhysRevD.34.1326}. }
    \label{fig:LLP_kinematics}
\end{figure}

This analysis is optimized for the LLP decays in the \hcal.
The kinematics of LLP production in a visibly decaying scenario are the same as those in an invisibly decaying scenario, in which the LLP carries away a significant fraction of the beam energy, leaving only a soft recoil electron measured in the tracker and \ecal.
Therefore, we expect the same physics signatures in the recoil tracker and \ecal as in the missing-momentum search, and we can employ many of the same event selection criteria for which LDMX was optimized~\cite{LDMX:2023zbn}, discussed in Section~\ref{subsec:trigger}.
The selection criteria for this analysis diverge from the missing-momentum search in the \hcal, where we expect a large amount of energy deposited from the LLP decay instead of no activity.
We expect the energy deposited in the \hcal to be near the beam energy and close to the beam axis, based on the LLP kinematics shown in Figure~\ref{fig:LLP_kinematics}.
Note that the angular distribution of LLPs is well within the minimum angular coverage of the Back \hcal (0.17 rad).

Sections \ref{subsec:hcalcuts} and \ref{subsec:bdt} describe our \hcal-specific event selection criteria, including a boosted decision tree (BDT), which is used to distinguish the electromagnetic shower of $A'$, $a$ $\rightarrow e^+e^-$ from hadronic showers originating from photo-nuclear-induced backgrounds.
As the production kinematics of $A'$ and ALP are similar and the signal signature of a displaced $e^+e^-$ pair is the same in both models, the same set of selection criteria is used.
The results of applying all event selection criteria on background and signal are summarized in Sections~\ref{subsec:totalbkg} and~\ref{subsec:signaleff}, respectively.

\subsection{Trigger and Missing-Momentum Event Selection}\label{subsec:trigger}

Events are triggered by less than \qty{3160}{\MeV} deposited in the first \num{20} layers of the \ecal (out of \num{34} total layers).
Following the trigger, three additional selection criteria are applied to reject backgrounds based on activity in the recoil tracker and \ecal. The first is a more stringent missing energy requirement in the entire \ecal, which requires less than \qty{3160}{\MeV} deposited in all \ecal layers---the same energy requirement as the first 20 \ecal layers from the trigger. Next, we require exactly one track reconstructed in the recoil tracker with a momentum less than \qty{2400}{\MeV}, to ensure that the recoil electron lost a significant amount of momentum in the target.
Any remaining background events will consist of events in which the electron loses a significant fraction of its energy to a bremsstrahlung photon that does not deposit all of its energy in the \ecal; this is characteristic of PN reactions or muon pair production.
The high granularity of the \ecal can be used to distinguish whether photon-induced energy depositions accompany the recoil electron shower, which should be the only \ecal feature in a signal event.
We use a BDT developed for the missing-momentum analysis to distinguish between signal and background events based on shower shape features and MIP tracks in the \ecal.
Details on the features implemented in the missing-momentum BDT can be found in~\cite{TDR}.
We use the same discriminator threshold for this BDT as in the missing-momentum search, which is set in conjunction with the veto requirements of the \hcal.

\subsection{\hcal Selection Criteria}\label{subsec:hcalcuts}

Up to this point, the same selection criteria as used in the missing-momentum search were applied. In this analysis, we utilize the HCal as a means of detecting a visibly decaying LLP.
A visibly decaying LLP will leave an energy deposition in the \hcal of the amount lost by the incident electron.
Therefore, to identify the signature of a visible decay, more than \qty{4840}{\MeV} must be deposited in the \hcal.
This complements the trigger requirement of less than \qty{3160}{\MeV} deposited in the \ecal.
Furthermore, to capture LLP decays that occur inside the \hcal, the start of the shower must be fully contained in the \hcal.
This is imposed by requiring that the reconstructed shower not start in the first layer of the \hcal, or that fewer than 5 photo-electrons (0.34\,MeV, the median noise level in the \hcal~\cite{TDR}) are deposited in any bar in the first layer of the \hcal.
This mitigates against backgrounds with activity upstream of the \hcal.

\subsection{\bdt}\label{subsec:bdt}

Background events that leave very little activity in the \ecal and a large amount of energy in the \hcal will originate from PN reactions producing neutral hadrons (such as $K_S, K_L, n$) and muon pair production.
These background events will primarily leave MIP tracks or hadronic showers in the \hcal, whereas the visible LLP decay signal will leave an electromagnetic shower.
To mitigate this remaining background, a BDT is utilized and is trained to distinguish between electromagnetic and hadronic showers in the \hcal.

This BDT (labeled ``\bdt'' to distinguish it from the BDT used in the missing-momentum search) is trained on \hcal shower shape features and other kinematic features to distinguish signal from background.
For the signal, we expect a single electromagnetic shower lying along the expected trajectory of a particle produced in the target.
For the background, we expect hadronic showers, MIP tracks, and generally higher shower multiplicity than in the signal.
The \bdt was implemented with XGBoost~\cite{xgboost} and utilizes twelve features to capture the differences we qualitatively expect from signal and background: the total number of \hcal layers hit; the energy-weighted hit $x$, $y$ and $z$ standard deviation (coordinate system shown in Figure~\ref{fig:ldmx_visibles_cartoon}); the energy-weighted hit $x$, $y$, and $r$ mean position ($r = \sqrt{x^2 + y^2}$); the number of isolated (defined later) hits; the isolated energy; the total number of \hcal hits; the total energy deposited in the \hcal; and the energy-weighted mean hit distance from the projected photon trajectory (defined later).
Shape distributions for these features, for both signal and ECal PN backgrounds, are shown in the left panel of Figure~\ref{fig:BDT_output} and in Appendix~\ref{app:hcal_bdt_feat}.

In the \hcal, a hit is defined to be ``isolated'' if there are no additional hits in a neighboring counter in the same layer. The isolated energy is the total energy of the isolated hits.
The energy-weighted mean and standard deviation of the hits in $x$, as an example, are defined as:
\begin{equation}
    \mu_x = \frac{\sum_{i}x_iE_i}{E_{\text{tot}}}, \qquad \sigma_x = \sqrt{\frac{\sum_{i}E_i(x_i-\mu_x)^2}{E_\text{tot}}}.
\end{equation}
Similar definitions follow for other energy-weighted features.

The distance of hits from the projected photon trajectory is a strong discriminator between signal and background.
For bremsstrahlung or LLP production in the target, there are two outgoing particles: a recoil electron and an unobserved photon or a LLP.
The recoil tracker measures the momentum of the recoil electron and, with the tagger tracker, the momentum of the unobserved particle can be determined.
The trajectory of the unobserved particle is then projected to the \hcal, and the average distance in the $(x,y)$ plane of the reconstructed hits from the trajectory is calculated.
Signal events from an LLP decay will have an electromagnetic shower centered on this trajectory, so we expect the mean distance of all hits from this trajectory to be small.
PN events will have larger distances from the trajectory due to wider hadronic showers or high particle multiplicities in the final state.
The distribution of this feature for signal and background is shown in Figure~\ref{fig:BDT_output}.

The \bdt was trained on \ecal PN background events and $A'$ events composed of an approximately equal mix of four $A'$ masses (5, 10, 50, and 100 MeV).
As seen in the BDT feature distributions in Appendix~\ref{app:hcal_bdt_feat}, $A'$ events and ALP events are very similar and clearly distinguishable from \ecal PN events.
For each simulated mass $\sim$65,000 events were used for training, along with a similar number of ECal PN events.
All events included in the training passed the trigger, tracker veto, and \hcal energy requirement.
Note that the \ecal BDT selection criterion is not enforced on the training sample to ensure a high-statistics sample for training.
Even if these events would ultimately be vetoed before the \bdt criterion due to their topology in the \ecal, they hold useful information for the \bdt on the topology of background events in the \hcal, and therefore, we include them in the training sample. 

The performance of the \bdt was assessed for separate batches of \ecal PN events and each signal mass and model individually, with all other analysis cuts applied.
The \bdt output scores for these test samples are shown in Figure~\ref{fig:BDT_output} for the \ecal PN background and two $A'$ and ALP masses.
Clear discrimination between signal and background is evident for all example masses and models.
Heavier mass LLPs tend to have a longer tail for low BDT scores and look more background-like than lighter masses.
This is because their electromagnetic showers are slightly longer and wider at higher masses, which more closely resemble the hadronic showers of the background.
The discriminator threshold for the \bdt is chosen based on a Figure of Merit (FOM) proposed by Punzi~\cite{punzi2003}.
At the FOM's optimal threshold, zero background events remain;
however, to ensure more robust background rejection and to take into account random fluctuations, a slightly higher, more conservative discriminator threshold is chosen that further reduces the signal efficiency by 5\%~\cite{horoho2025}.
This decision acknowledges the inherent variability of the background sampling in our simulation, but is at the expense of signal efficiency.
The choice of discriminator threshold also provides further background rejection for larger sample sizes like the targeted \num{1e16} EoT.

 \begin{figure}
    \centering
    \includegraphics[width=0.45\linewidth]
    {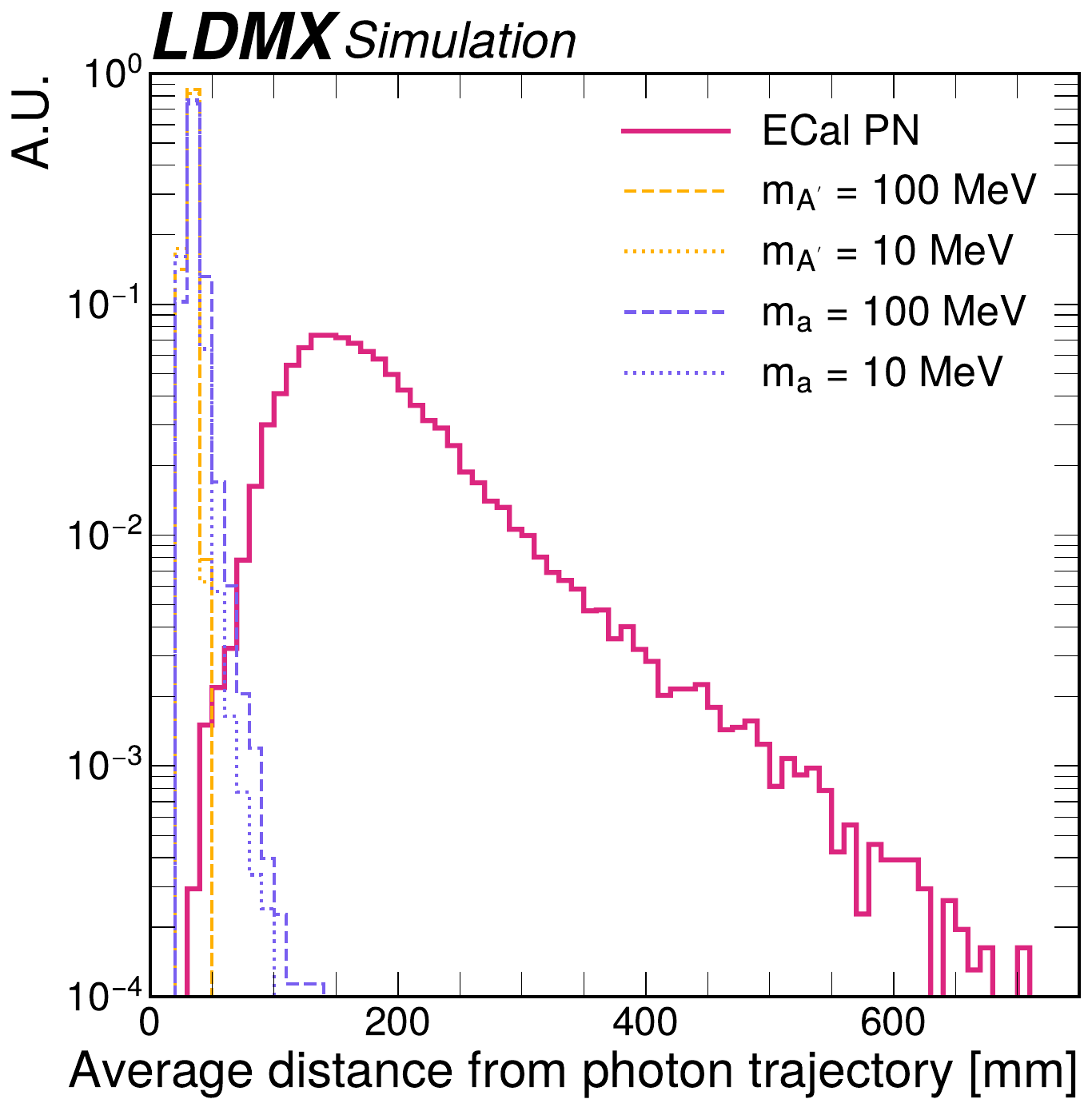} 
    \includegraphics[width=0.45\linewidth]{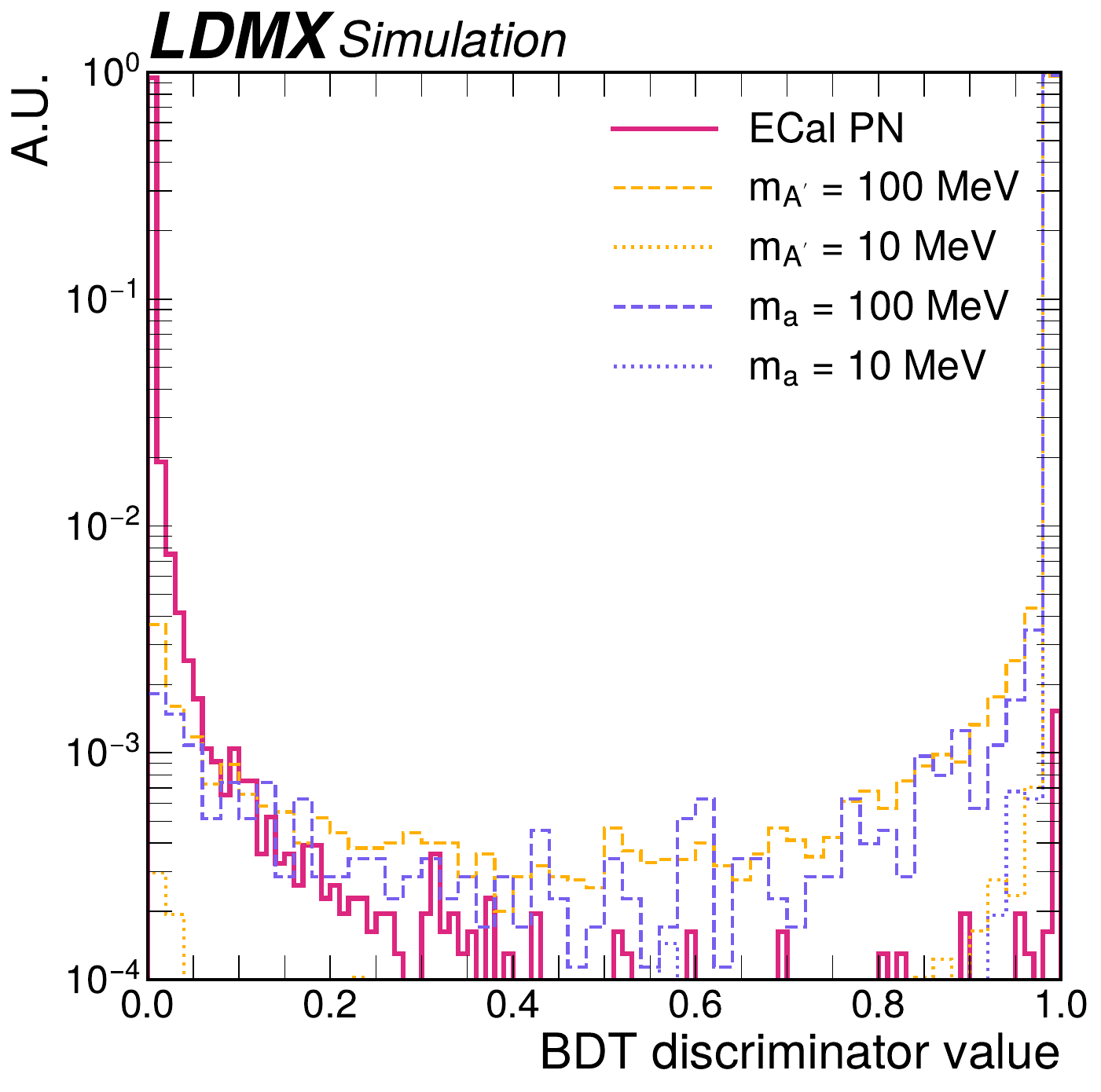} 
    \caption{Left: The signal and background distributions for one of the top performing BDT features, the average distance of \hcal hits from the projected photon trajectory. Right: The \bdt discriminator value output. Events in these plots have all cuts applied except the \bdt.}
    \label{fig:BDT_output}
 \end{figure}

\subsection{Total Background}\label{subsec:totalbkg}

The set of selection criteria outlined in the earlier sections are tested on the simulated PN and muon photoproduction events samples discussed in Sec.~\ref{subsec:background_simulation}.
The sizes of these samples range from \num{e14} - \num{e15} EoT.
Table~\ref{tab:background_cuts} shows how the expected yields of each background change as each selection criterion is applied.
The most challenging backgrounds to veto come from single, neutral kaons produced in PN reactions in the \ecal and decaying to $\pi^0$ mesons in the \hcal.
These events appear signal-like due to their electromagnetic final states.
Our requirement on the \bdt discriminator value vetoes these remaining events.
Muon photoproduction backgrounds are easy to veto due to the presence of MIP tracks, with none remaining following the requirement on ECal BDT score.
For all backgrounds, zero events remain, and the presented analysis is capable of rejecting all photon-induced backgrounds at \num{e14} EoT.
In Sec.~\ref{sec:uncertainty}, we discuss our current dependence on simulation and summarize studies that were carried out to quantify the effect of hadronic modeling uncertainties.

\begin{center}
\begin{table}[h]
    \centering
    \caption{Number of simulated background events remaining at different selection stages. The background classes are separated by where in the detector the reaction took place.}
    \label{tab:background_cuts}
    \resizebox{\textwidth}{!}{
    \begin{tabular}{l|c|c|c|c}
        \hline
        \hline
        \multirow{2}{*}{ } & \multicolumn{2}{c|}{\textbf{Photo-nuclear}} & \multicolumn{2}{c}{\textbf{Muon photoproduction}} \\
        \cline{2-5}
        & \textbf{Target} & \textbf{ECal} & \textbf{Target} & \textbf{ECal}\\
        \hline
        EoT equivalent & $1.03\times10^{14}$ & $1.00\times10^{14}$ & $1.00\times10^{15}$ & $1.00\times10^{14}$ \\
        \hline
        \hline
        Trigger (front ECal energy < 3160\,MeV) & $1.77\times10^7$ & $1.61\times10^8$ & $4.56\times10^6$ & $1.47\times10^7$ \\
        Total ECal energy < 3160\,MeV & $1.22\times10^7$ & $3.87\times10^7$ & $3.46\times10^6$ & $1.02\times10^7$ \\
        Single track with p < 2400\,MeV & 69282 & $3.50\times10^7$ & 16101 & $9.86\times10^6$ \\
        Missing-momentum ECal BDT & 1406 & 61145 & $< 1$ & $< 1$ \\
        Main HCal energy > 4840\,MeV & 755 & 55283 & $< 1$ & $< 1$ \\
        HCal Containment requirement & 251 & 30655 & $< 1$ & $<1$ \\
        Visibles BDT  & $< 1$ & $<1$ & $<1$ & $<1$ \\
        \hline
        \hline
    \end{tabular}
    }
\end{table}
\end{center}

\subsection{Signal Efficiency}\label{subsec:signaleff}

The effectiveness of the detailed selection criteria in identifying signal events is summarized by the signal efficiencies presented in Table~\ref{tab:signal_cuts}, and the final signal efficiency as a function of distance from the target is shown in Figure~\ref{fig:signaleff}.
The selection cuts become efficient to the signal at 900\,mm from the target, which is due to the requirement of no activity in the first layer of the \hcal.
We find that the signal efficiency is approximately constant along the distance of the \hcal, and therefore the efficiency values reported in the table are the average efficiency in the first 1\,m of the \hcal.
The signal efficiency has a slight dependence on the mass of the LLP, with heavier masses having a worse efficiency.
This is due to the \bdt's worsened performance on higher mass LLPs, explained above.
After the application of all the selection criteria, the observed efficiencies for all tested $A'$ and ALP masses are in the range 27 - 38\,\%.

\begin{center}
\begin{table}[h]
    \centering
    \caption{The signal efficiency in the \hcal at different selection stages for two $A'$ and ALP masses. The efficiency is measured as the average efficiency in the first 1\,m of the \hcal.}
    \resizebox{\textwidth}{!}{
    \begin{tabular}{l|c|c|c|c}
        \hline
        \hline
        \multirow{2}{*}{ } & \multicolumn{2}{c|}{\textbf{$A'$ efficiency (\%)}} & \multicolumn{2}{c}{\textbf{ALP efficiency (\%)}}  \\
        \cline{2-5}
        & \textbf{10 MeV} & \textbf{100 MeV} & \textbf{10 MeV} & \textbf{100 MeV} \\
        \hline
        Tracker acceptance (p>50 MeV) & 90.8 & 77.2 & 76.5 & 65.7\\
        Trigger (front ECal energy < 3160\,MeV) &  75.4 & 69.1  & 74.2 & 64.5 \\
        Total ECal energy < 3160\,MeV & 75.2 & 69.0 & 74.2& 64.4\\
        Single track with p < 2400\,MeV & 69.1 & 65.0  & 72.4 & 61.6\\
        Missing-momentum ECal BDT & 64.1 & 62.5 & 69.5 & 61.1\\
        Main HCal energy > 4840\,MeV & 61.7 & 58.9 & 68.1& 59.4\\
        Containment requirement & 61.7 & 58.9  & 68.0 & 59.2\\
        Visibles BDT & 33.4 & 27.0 & 38.5 & 27.7\\
        \hline
        \hline
    \end{tabular}
    }
    \label{tab:signal_cuts}
\end{table}
\end{center}

The large majority of events pass the fiducial acceptance requirement (a recoil electron with $p>50$ MeV), although there are more soft recoil electrons accompanying the ALP than the $A'$.
Higher mass LLPs are more likely to produce softer recoil electrons; thus, the efficiency for the 100 MeV is worse than that for the 10 MeV case.
The most significant source of efficiency loss results from the conservative Visibles BDT threshold.
This single criterion removes about half of the remaining events, and all preceding criteria collectively remove about 40 $\%$ of the initial events.
This substantial efficiency loss is required to reduce the yield of background events at $1\times10^{14}$ EoT from about 31,000 to less than one event.

\section{Systematic Uncertainties \label{sec:uncertainty}}

This analysis has shown that LDMX can achieve a background-free search at $10^{14}$ EoT and a minimum 27\% signal efficiency.
However, there are several uncertainties in the presented sensitivity study, discussed below.

Before embarking on physics analysis, the LDMX collaboration will conduct a comprehensive commissioning phase that includes detector calibration and in-situ background validation. This ensures that the final physics searches will use a well-understood, data-validated MC model. This robust validation process ensures that the simulation uncertainties present in preliminary studies will not compromise the final physics reach, and it is expected that the data-validated BDT will provide similar or improved background discrimination capabilities.

\subsection{Shower Shape Uncertainties}

This analysis relies on simulations of electromagnetic and hadronic showers that are central to our primary background rejection technique: using shower shape features in a BDT.
The effectiveness of this BDT is therefore directly dependent on the chosen {\sc Geant4} physics model~\cite{geant4_2016}, with the default list in \ldmxsw being FTFP-BERT, which combines the Fritiof (FTF) string model for high-energy interactions with a custom Bertini (BERT) intra-nuclear cascade model for lower energies, detailed in \cite{LDMX:2018cma}.

To assess the systematic uncertainty related to the physics list choice, we simulated photonuclear background processes using five alternative physics lists to understand the impact on shower shapes and secondary particle production. Overall the largest change in the mean and RMS of the BDT input features was $\mathcal{O}(10\%)$, but background shapes remained distinctly different from the signal shape. To quantify the effect of these uncertainties on our final result, we evaluated the change in BDT scores when using these updated feature distributions. We found that the optimal BDT cut still provides very good background discrimination and in all cases no background
events passed the signal selection criteria. %We found that the optimal BDT cut still provides very good background discrimination, in all cases we continued to have no background events passing the signal selection criteria.

\subsection{Theoretical Uncertainties}

The normalization of our signal rates is taken from the equations of Ref.~\cite{Berlin:2018bsc}.
Underlying these equations are derived cross sections for the production and decay modes and these have associated uncertainties.
The modeling of the virtual photon-nucleus interaction within {\sc MadGraph/MadEvent} is done using standard atomic/nuclear form factors and there is a 4$\%$ uncertainty on the calculated cross section~\cite{RevModPhys.46.815} resulting from modeling uncertainties.
This level of uncertainty does not produce a shift in our maximal physics reach.
Statistical uncertainties have a negligible effect on the final result.  

\subsection{Detector/Reconstruction Uncertainties}

\subsubsection{Tracker Uncertainty}
One of the most powerful \bdt features for discriminating signal and background is the energy-weighted mean hit distance from the projected photon trajectory, which relies on an accurate measurement of the recoil electron momentum to calculate the projected photon path.
For recoil electron tracks with momenta below 2400\,MeV, the $p_T$ uncertainty will be no more than 3.4\,MeV~\cite{TDR}.
The effect of this uncertainty on the \bdt performance was studied and found to have a 1\% decrease in the signal efficiency and induced an uncertainty of one background event for \bdt discriminator values near the threshold.
However, there are still no background events in the signal region after introducing uncertainty due to tracker resolution.

\subsubsection{HCal Uncertainty}

This analysis relies on the HCal's ability to distinguish electromagnetic showers of LLP decays from hadronic showers and MIP background tracks produced by SM backgrounds.
The HCal's intrinsic resolution is discussed in \cite{TDR}.
To test the robustness of our analysis against imperfect energy resolution modeling in the current simulation, we simulated a 5$\%$ resolution uncertainty.
To simulate imperfect resolution modeling, an additional Gaussian smearing ($\mu = 1$, $\sigma$ = 0.05) is added to the reconstructed HCal energy sum.
Our BDT, trained on the unsmeared Monte Carlo samples, still achieved good signal-background separation when applied to these smeared samples, confirming that the analysis strategy is robust against detector resolution effects. 
Light yield decline due to scintillator aging is also not expected to affect the performance of this analysis.
The aging of similar counters for the Mu2e cosmic ray veto system were measured to be $\sim$3\%/yr~\cite{corrodi2026}.

\subsubsection{ECal Uncertainty}

Although our primary mode of detection is the HCal, we utilize the ECal as a means of triggering our event, and ECal shower features form part of our selection procedure through the ECal BDT criterion.
The overall calibration requirements on the ECal are for 5$\%$ inter-module calibration and 1$\%$ scale, neither had any observable effect on total background events passing all selection criteria in this analysis.

\section{Projected Sensitivity}\label{sec:ratesandreach}

\subsection{Statistical Treatment}

In the following section, our projected physics reach for both signal scenarios is presented.
The $CL_s$ treatment~\cite{CLs_1999,CLs_2002} is used, which introduces a modified test statistic, ensuring that the limit does not over-exclude hypotheses in regions of low sensitivity.
This is crucial for searches involving rare events where the number of expected signal events is small (as is the case for the presented searches).
The $CL_s$ statistical treatment is widely adopted in high-energy physics to set conservative upper limits on non-negative parameters.

For the two analyses presented here, we take the number of expected background events at \num{4e14} EoT to be 0.5 and assume 0 observed events.
A Poisson distribution is assumed for both the expected number of signal and background events.
Based on the uncertainties discussed in Sec.~\ref{sec:uncertainty}, a conservative 10\% uncertainty in the signal is used when calculating the number of signal events at 90\% confidence.
For the background, statistical uncertainties are dominant.
Although the presented analysis finds no background events passing the selection criteria, due to computational constraints, the simulated background samples were equivalent to $1\times10^{14}$ EoT.
To account for the limited size of our simulated samples when extrapolating our sensitivity to \num{1e16} EoT, we present three scenarios: 0.5 expected background events and 0 observed events, 5 expected background events and 5 observed events, and 50 expected background events and 50 observed events.

\subsection{Sensitivity Estimates}

Reach estimates for the minimal $A'$ and ALP$-e$ models can be constructed from theoretical calculations of signal production and lifetime detailed in~\cite{Berlin:2018bsc} along with knowledge of the expected signal efficiency and background estimates detailed in this work.
A few conservative assumptions are made to simplify the calculation of the reach.
First, as seen in Figure~\ref{fig:LLP_kinematics}, the fraction of beam energy carried away by the ALP or $A'$ differs between masses and is not monoenergetic.
We make a conservative approximation $E_{a}/E_{\text{beam}} = 0.93$ and $E_{A'}/E_{\text{beam}} = 0.75$, which are the lowest average energy fractions for the ALP and $A'$ masses tested.
These approximations are used for all ALP and $A'$ masses.
On average ALPs produce slightly softer recoil electrons, and themselves take larger fractions of the beam energy.
Second, the kinematics of both ALP and $A'$ production are such that they are very forward with respect to the $z$ (horizontal) axis, seen in Figure~\ref{fig:LLP_kinematics}.
Therefore, the total distance traveled by the LLP can be approximated as the distance from the target along the beam axis.
A minimum travel distance of \qty{950}{\mm} is assumed, corresponding to the distance from the target at which the analysis cuts have constant efficiency, as well as a constant signal efficiency of 27\% for all $A'$ and ALP masses.
These assumptions are justified by the fact that the efficiency for a given signal sample is constant throughout the HCal (see Appendix ~\ref{app:sig_eff_plot}).

\begin{figure}[htpb]
    \centering
    \includegraphics[width=0.95\linewidth]{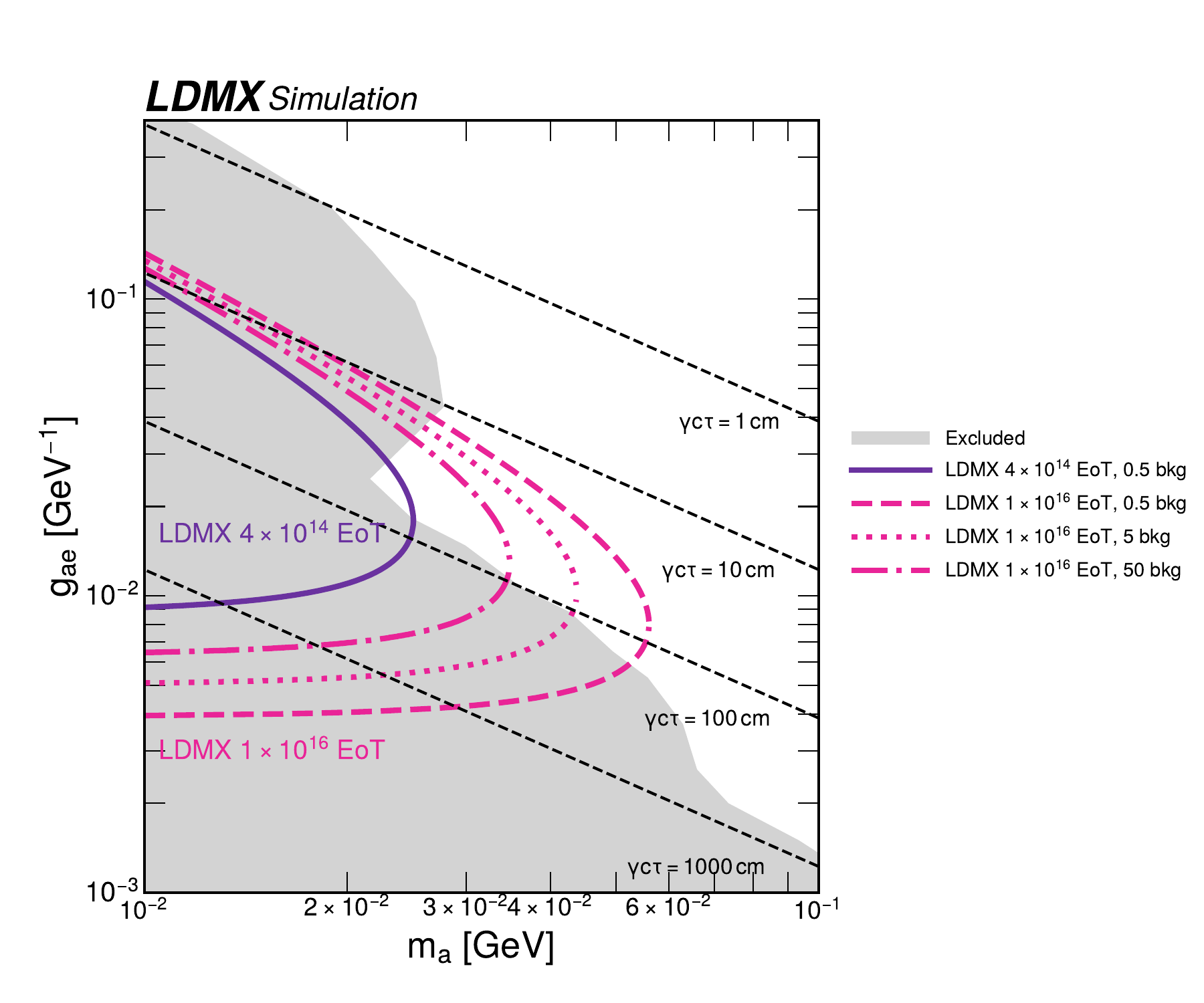}
    \caption{\CL sensitivity to ALP-$e$ for LDMX projected for \num{4e14} EoT and \num{e16} EoT. Existing limits from previous experiments~\cite{Riordan:1987aw, Bjorken:1988as, Davier:1989wz} are shown in the gray shaded region. The solid curve assumes a mean background of 0.5 events. The dashed curves assume different mean backgrounds at $1 \times 10^{16}$ EoT: 0.5 background events (dashed), 5 background events (dotted), and 50 background events (dash-dotted).}
    \label{fig:final_reach_alps_elec}
\end{figure}

\begin{figure}[htpb]
    \centering
    \includegraphics[width=0.95\linewidth]{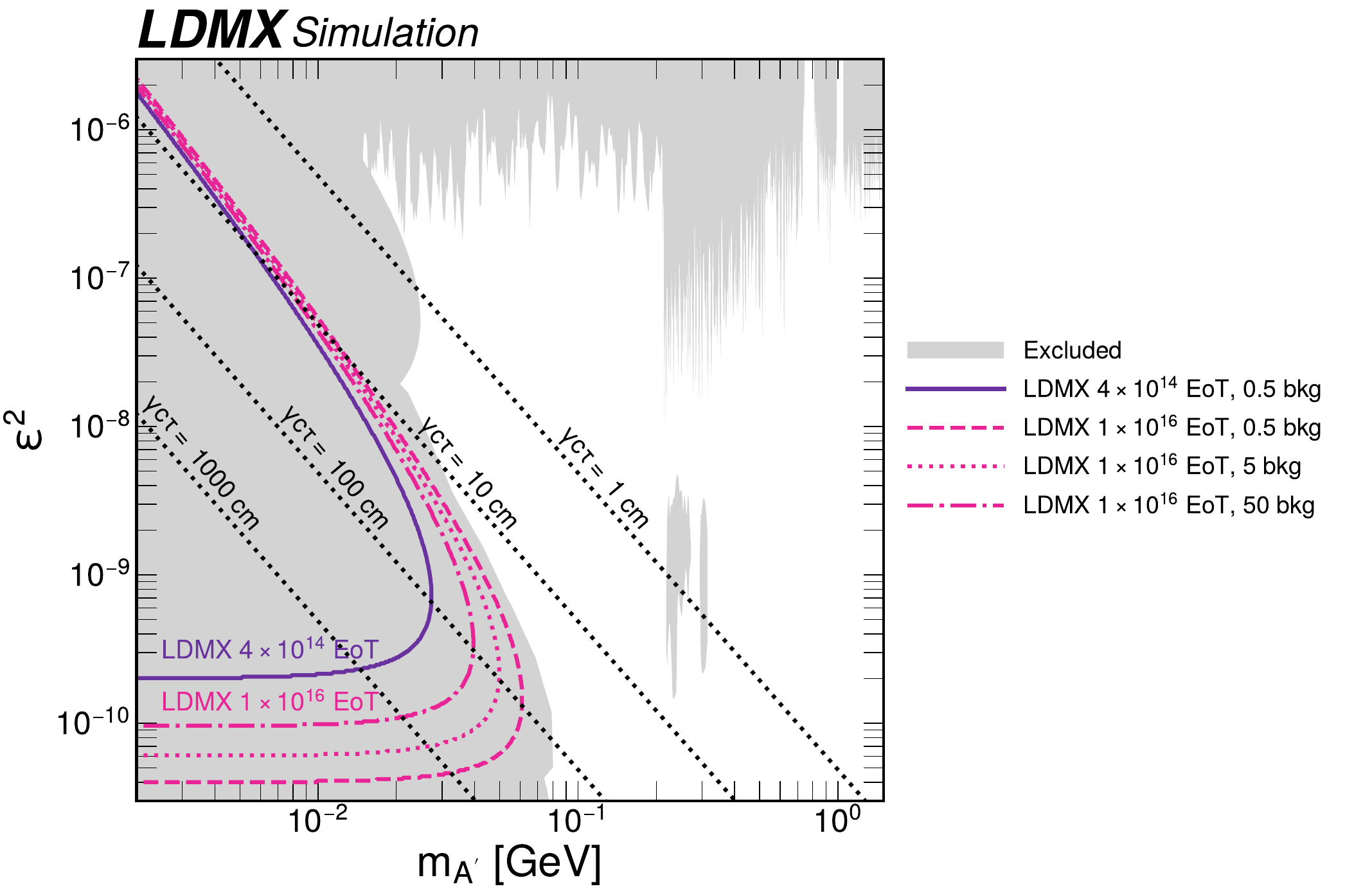}
    \caption{\CL sensitivity to $A'$ for LDMX projected for \num{4e14} EoT and \num{e16} EoT. Existing limits from previous experiments~\cite{BABAR_2009, KLOE_2012, KLOE_2013, LHCb_2020, FASER_2023, KEK_Orsay_update_2012, U70_1991, U70_1992, NA64_2020, NA48_2_2015} are shown in the gray shaded region. The solid curve assumes a mean background of 0.5 events. The dashed curves assume different mean backgrounds at \num{1e16} EoT: 0.5 background events (dashed), 5 background events (dotted), and 50 background events (dash-dotted).}
    \label{fig:final_reach_aprime}
\end{figure}

Reach estimates are calculated at \qty{4e14}{EoT} with 0.5 background events and \qty{1e16}{EoT} with 0.5 background events, 5 background events, and 50 background events.
Figure~\ref{fig:final_reach_alps_elec} shows the projected physics reach for the ALP-$e$ search at LDMX in these scenarios, compared to existing experimental limits~\cite{Riordan:1987aw, Davier:1989wz, Bjorken:1988as}.
Figure~\ref{fig:final_reach_aprime} shows the projected physics reach for the $A'$ search at LDMX, compared to existing experimental limits~\cite{BABAR_2009, KLOE_2012, KLOE_2013, LHCb_2020, FASER_2023, KEK_Orsay_update_2012, U70_1991, U70_1992, NA64_2020, NA48_2_2015}.

\section{Future Targets and Improvements}\label{sec:improvements}

The presented analysis strategy provides good signal efficiency and background rejection.
However, we acknowledge that in the future the sensitivity of LDMX to $A'$ or ALPs could be further extended, providing additional unique physics reach.

Regarding the selection criteria, the \ecal shower shape veto used in the missing-momentum analysis can be improved and updated.
The discriminator threshold for the BDT was set based on the requirements of the missing-momentum search. Optimization of the missing-momentum BDT threshold in conjunction with the \bdt threshold may provide better background rejection or higher signal efficiency.
Furthermore, other veto strategies for \ecal shower topology are possible. Preliminary studies on the use of a graph neural network, like ParticleNet~\cite{qu2020}, show that this type of machine learning model may be more effective as a veto than a BDT~\cite{TDR}.
The high granularity of the ECal, with more than 100,000 readout channels, is more effectively leveraged with a neural network than a BDT.
Since this analysis is background-free with a BDT at \num{1e14}\,EoT, it is likely that a neural network can keep the analysis background-free at the same or larger EoT. 

In addition, this analysis focuses on LLPs produced at the target, which must travel $\sim$1\,m before decaying to be fully contained in the \hcal.
The sensitivity of LDMX to visibly decaying LLP signatures could be extended to shorter lifetimes by using the \ecal as a target~\cite{EaT} while continuing to utilize the \hcal as the primary detector.
In addition, the production of LLPs through the decay of neutral mesons in the \ecal would increase the yield of signal events.
An auxiliary search such as this one strengthens the broader physics case for LDMX.

Another method to improve LDMX's sensitivity to shorter lifetimes is through a recoil tracker search for ``dark tridents'' (the recoil electron together with an $e^+e^-$ pair from a LLP decay)~\cite{gaiser2026}.
This would require a new trigger and a very different analysis approach from what is presented here, which will be the subject of future work.

When the ALP-$\gamma$ coupling dominates the kinematics of the outgoing ALP are very different. The invisible ALP takes a much smaller fraction of the beam energy, consequently, the majority of events fail the current missing energy trigger. A custom HCAL trigger is currently under development to achieve an efficiency comparable to that of the ALP-$e$ channel. Due to the distinct trigger scheme and the differing set of backgrounds requiring mitigation, this analysis will be discussed in detail in a future publication.

\section{Conclusions}\label{sec:conclusion}

LDMX is a proposed electron-beam fixed-target experiment primarily designed for a world-leading, model-independent, search for sub-GeV dark matter using the missing-momentum technique. Beyond this primary goal, the detector design of LDMX is well-suited for probing other dark sector phenomena, specifically those involving long-lived, visibly decaying particles.

In this article, we present the first comprehensive detector simulation and analysis strategy for searching for the visible decay signatures of minimal $A'$ and ALP using an \qty{8}{\GeV} electron beam in LDMX.
Our results show a competitive physics reach for these LLP signatures, as illustrated by the projected 90$\%$ C.L. sensitivity curves in Figures \ref{fig:final_reach_aprime} and \ref{fig:final_reach_alps_elec}.
The presented simulations incorporate realistic detection efficiencies and predicted background levels, demonstrating sensitivity comparable to other ongoing experiments.
For the ALP search, our proposed analysis provides unique sensitivity in the mass region from $\sim$20 to 56 MeV for \num{e16} EoT, which is beyond the current capabilities of other current/upcoming experiments.%, including the FASER experiment, for this particular model.

The physics models investigated in this work are distinct from the flagship missing-momentum analysis, yet are complementary to LDMX's overall program.
We establish that LDMX can achieve a unique physics reach for LLPs that produce electromagnetic signatures within the \hcal.
The robustness of the analysis procedure stems from its ability to effectively separate electromagnetic signals from hadronic backgrounds.
Although ALP and $A'$ models serve as benchmarks, this analysis strategy can be adapted to any LLP model that produces similar, high-energy electromagnetic, visible signatures in the LDMX detector.

Ultimately, by conducting searches for both invisible dark matter and visibly decaying LLPs, LDMX will make a significant contribution to the search for light dark matter and provide a broad exploration of the sub-GeV dark sector.

\acknowledgments

We would like to thank Nikita Blinov for providing the generator used for the ALP simulation and for answering our questions during this analysis.

Support for UCSB is made possible by the Joe and Pat Yzurdiaga endowed chair in experimental science. 
Use was made of the UCSB computational facilities administered by the Center for Scientific Computing at the California NanoSystems Institute and Materials Research Laboratory (an NSF MRSEC; DMR-1720256) and purchased through NSF CNS-1725797, and from resources provided by the Swedish National Infrastructure for Computing at the Centre for Scientific and Technical Computing at Lund University (LUNARC), as well as LUNARC’s own infrastructure. Contributions from Caltech, CMU, Stanford, UMN, UVA, and UCSB are supported by the US Department of Energy under grants DE-SC0011925, DE-SC0010118, DE-SC0022083, DE-SC00012069, DE-SC0007838, and DE-SC0011702, respectively. Participation by Lund University is enabled by the Knut and Alice Wallenberg Foundation (Dnr. KAW 2018.0429, KAW 2019.0080, KAW 2023.0064). RP acknowledges support from the Swedish Research Council (Dnr 2019-03436). AW, GK, WK, and NT are supported by Fermi Forward Discovery Group, LLC, acting under Contract No. 89243024CSC000002 with the U.S. Department of Energy, Office of Science, Office of High Energy Physics, and the Fermilab LDRD program.  EP, MG, TN, PS, and NT are supported by Stanford University under Contract No. DE-AC02-76SF00515 with the U.S. Department of Energy, Office of Science, Office of High Energy Physics.

% Bibliography

%% [A] Recommended: using JHEP.bst file
\bibliographystyle{JHEP}
\bibliography{references.bib}

@article{LDMX:2023zbn,
    author = "\r{A}kesson, Torsten and others",
    collaboration = "LDMX",
    title = "{Photon-rejection power of the Light Dark Matter eXperiment in an 8 GeV beam}",
    reportNumber = "FERMILAB-PUB-23-433-PPD-T, SLAC-PUB-17550",
    doi = "10.1007/JHEP12(2023)092",
    journal = "JHEP",
    volume = "12",
    pages = "092",
    year = "2023"
}

@article{EaT,
    author = "{\r{A}}kesson, Torsten and others",
    collaboration = "LDMX",
    title = "{Sensitivity of an Early Dark Matter Search using the Electromagnetic Calorimeter as a Target for the Light Dark Matter eXperiment}",
    reportNumber = "FERMILAB-PUB-25-0583-CSAID-PPD",
    doi = "https://doi.org/10.1007/JHEP12(2025)150",
    journal = "J. High Energ. Phys.",
    volume = "2025",
    pages="150",
    year = "2025"
}

@article{TDR,
    author = "Appert, Stephen and others",
    collaboration = "LDMX",
    title = "{LDMX -- The Light Dark Matter eXperiment}",
    eprint = "2508.11833",
    archivePrefix = "arXiv",
    primaryClass = "hep-ex",
    reportNumber = "FERMILAB-PUB-25-0605-CSAID-PPD",
    month = "8",
    year = "2025"
}

@article{Riordan:1987aw,
    author = "Riordan, E. M. and others",
    title = "{A Search for Short Lived Axions in an Electron Beam Dump Experiment}",
    reportNumber = "SLAC-PUB-4280, UR-993, FERMILAB-PUB-87-251",
    doi = "10.1103/PhysRevLett.59.755",
    journal = "Phys. Rev. Lett.",
    volume = "59",
    pages = "755",
    year = "1987"
}

@article{Bjorken:1988as,
    author = "Bjorken, J. D. and Ecklund, S. and Nelson, W. R. and Abashian, A. and Church, C. and Lu, B. and Mo, L. W. and Nunamaker, T. A. and Rassmann, P.",
    title = "{Search for Neutral Metastable Penetrating Particles Produced in the SLAC Beam Dump}",
    reportNumber = "FERMILAB-PUB-88-044, PRINT-88-0352 (FERMILAB)",
    doi = "10.1103/PhysRevD.38.3375",
    journal = "Phys. Rev. D",
    volume = "38",
    pages = "3375",
    year = "1988"
}

@article{Davier:1989wz,
    author = "Davier, M. and Nguyen Ngoc, H.",
    title = "{An Unambiguous Search for a Light Higgs Boson}",
    reportNumber = "LAL 89-24",
    doi = "10.1016/0370-2693(89)90174-3",
    journal = "Phys. Lett. B",
    volume = "229",
    pages = "150--155",
    year = "1989"
}

@inproceedings{Raubenheimer:2018wwc,
    author = "Raubenheimer, Tor",
    title = "{The LCLS-II-HE, A High Energy Upgrade of the LCLS-II}",
    booktitle = "{60th ICFA Advanced Beam Dynamics Workshop on Future Light Sources}",
    doi = "10.18429/JACoW-FLS2018-MOP1WA02",
    pages = "MOP1WA02",
    year = "2018"
}

@article{LDMX:2018cma,
    author = "\r{A}kesson, Torsten and others",
    collaboration = "LDMX",
    title = "{Light Dark Matter eXperiment (LDMX)}",
    eprint = "1808.05219",
    archivePrefix = "arXiv",
    primaryClass = "hep-ex",
    reportNumber = "FERMILAB-PUB-18-324-A, SLAC-PUB-17303",
    month = "8",
    year = "2018"
}

@article{Berlin:2018bsc,
    author = "Berlin, Asher and Blinov, Nikita and Krnjaic, Gordan and Schuster, Philip and Toro, Natalia",
    title = "{Dark Matter, Millicharges, Axion and Scalar Particles, Gauge Bosons, and Other New Physics with LDMX}",
    reportNumber = "FERMILAB-PUB-18-310-A, SLAC-PUB-17297",
    doi = "10.1103/PhysRevD.99.075001",
    journal = "Phys. Rev. D",
    volume = "99",
    number = "7",
    pages = "075001",
    year = "2019"
}

@article{FASER_2023,
title = {Search for dark photons with the FASER detector at the LHC},
journal = {Physics Letters B},
volume = {848},
pages = {138378},
year = {2024},
issn = {0370-2693},
doi = {https://doi.org/10.1016/j.physletb.2023.138378},
url = {https://www.sciencedirect.com/science/article/pii/S0370269323007128},
author = {Henso Abreu and others},
collaboration={FASER}
}

@article{BABAR_2009,
  title = {Search for Dimuon Decays of a Light Scalar Boson in Radiative Transitions $\ensuremath{\Upsilon}\ensuremath{\rightarrow}\ensuremath{\gamma}{A}^{0}$},
  author = {Aubert, B. and others},
  collaboration = {BABAR},
  journal = {Phys. Rev. Lett.},
  volume = {103},
  issue = {8},
  pages = {081803},
  numpages = {7},
  year = {2009},
  month = {Aug},
  publisher = {American Physical Society},
  doi = {10.1103/PhysRevLett.103.081803},
  url = {https://link.aps.org/doi/10.1103/PhysRevLett.103.081803}
}

@article{KLOE_2013,
   title = {Limit on the production of a light vector gauge boson in \ensuremath{\phi} meson decays with the KLOE detector},
   volume = {720},
   ISSN = {0370-2693},
   url = {http://dx.doi.org/10.1016/j.physletb.2013.01.067},
   DOI = {10.1016/j.physletb.2013.01.067},
   number = {1–3},
   journal = {Physics Letters B},
   publisher = {Elsevier BV},
   author = {Babusci, D. and others},
   collaboration = {KLOE-2}, 
   year = {2013},
   month = {Mar}, 
   pages = {111–115} 
}

@article{KLOE_2012,
   title = {Search for a vector gauge boson in \ensuremath{\phi} meson decays with the KLOE detector},
   volume = {706},
   ISSN = {0370-2693},
   url = {http://dx.doi.org/10.1016/j.physletb.2011.11.033},
   DOI = {10.1016/j.physletb.2011.11.033},
   number = {4–5},
   journal = {Physics Letters B},
   publisher = {Elsevier BV},
   author = {Archilli, F. and others},
   collaboration = {KLOE-2},
   year = {2012},
   month = {Jan}, 
   pages = {251–255} 
}

@article{LHCb_2020,
  title = {Search for ${A}^{\ensuremath{'}}\ensuremath{\rightarrow}{\ensuremath{\mu}}^{+}{\ensuremath{\mu}}^{\ensuremath{-}}$ Decays},
  author = {Aaij, R. and others},
  collaboration = {LHCb},
  journal = {Phys. Rev. Lett.},
  volume = {124},
  issue = {4},
  pages = {041801},
  numpages = {12},
  year = {2020},
  month = {Jan},
  publisher = {American Physical Society},
  doi = {10.1103/PhysRevLett.124.041801},
  url = {https://link.aps.org/doi/10.1103/PhysRevLett.124.041801}
}

@article{KEK_Orsay_update_2012,
  title = {New limits on hidden photons from past electron beam dumps},
  author = {Andreas, Sarah and Niebuhr, Carsten and Ringwald, Andreas},
  journal = {Phys. Rev. D},
  volume = {86},
  issue = {9},
  pages = {095019},
  numpages = {6},
  year = {2012},
  month = {Nov},
  publisher = {American Physical Society},
  doi = {10.1103/PhysRevD.86.095019},
  url = {https://link.aps.org/doi/10.1103/PhysRevD.86.095019}
}

@article{U70_1991,
    author = "Blumlein, J. and others",
    title = "{Limits on neutral light scalar and pseudoscalar particles in a proton beam dump experiment}",
    reportNumber = "PHE-90-03",
    doi = "10.1007/BF01548556",
    journal = "Z. Phys. C",
    volume = "51",
    pages = "341--350",
    year = "1991"
}

@article{U70_1992,
    author = "Blumlein, J. and others",
    title = "{Limits on the mass of light (pseudo)scalar particles from Bethe-Heitler e+ e- and mu+ mu- pair production in a proton - iron beam dump experiment}",
    reportNumber = "PHE-91-11",
    doi = "10.1142/S0217751X9200171X",
    journal = "Int. J. Mod. Phys. A",
    volume = "7",
    pages = "3835--3850",
    year = "1992"
}

@article{NA64_2020,
   title = {Improved limits on a hypothetical \ensuremath{X(16.7)} boson and a dark photon decaying into \ensuremath{e^{+}e^{-}} pairs},
   volume = {101},
   ISSN = {2470-0029},
   url = {http://dx.doi.org/10.1103/PhysRevD.101.071101},
   DOI = {10.1103/physrevd.101.071101},
   number = {7},
   journal = {Physical Review D},
   publisher = {American Physical Society (APS)},
   author = {Banerjee, D. and others},
   collaboration = {NA64},
   year = {2020},
   month = {Apr} 
}

@misc{NA48_2_2015,
    title = {Search for the dark photon in $\pi^0$ decays}, 
    author = {Batley, J. R. and others},
    collaboration = {NA48/2},
    year = {2015},
    eprint = {1504.00607},
    archivePrefix = {arXiv},
    primaryClass = {hep-ex},
    url = {https://arxiv.org/abs/1504.00607}, 
}

@article{CLs_1999,
   title = {Confidence level computation for combining searches with small statistics},
   volume = {434},
   ISSN = {0168-9002},
   url = {http://dx.doi.org/10.1016/S0168-9002(99)00498-2},
   DOI = {10.1016/s0168-9002(99)00498-2},
   number = {2–3},
   journal = {Nuclear Instruments and Methods in Physics Research Section A: Accelerators, Spectrometers, Detectors and Associated Equipment},
   publisher = {Elsevier BV},
   author = {Junk, Thomas},
   year = {1999},
   month = {Sep}, 
   pages={435–443} 
}

@article{CLs_2002,
    author = "Read, Alexander L.",
    editor = "Whalley, M. R. and Lyons, L.",
    title = "{Presentation of search results: The $CL_s$ technique}",
    doi = "10.1088/0954-3899/28/10/313",
    journal = "J. Phys. G",
    volume = "28",
    pages = "2693--2704",
    year = "2002"
}

@article{RevModPhys.46.815,
  title = {Pair production and bremsstrahlung of charged leptons},
  author = {Tsai, Yung-Su},
  journal = {Rev. Mod. Phys.},
  volume = {46},
  issue = {4},
  pages = {815--851},
  numpages = {0},
  year = {1974},
  month = {Oct},
  publisher = {American Physical Society},
  doi = {10.1103/RevModPhys.46.815},
  url = {https://link.aps.org/doi/10.1103/RevModPhys.46.815}
}

@article{PhysRevD.34.1326,
  title = {Axion bremsstrahlung by an electron beam},
  author = {Tsai, Yung Su},
  journal = {Phys. Rev. D},
  volume = {34},
  issue = {5},
  pages = {1326--1331},
  numpages = {0},
  year = {1986},
  month = {Sep},
  publisher = {American Physical Society},
  doi = {10.1103/PhysRevD.34.1326},
  url = {https://link.aps.org/doi/10.1103/PhysRevD.34.1326}
}

@article{punzi2003,
    author = "Punzi, Giovanni",
    editor = "Lyons, L. and Mount, R. P. and Reitmeyer, R.",
    title = "{Sensitivity of searches for new signals and its optimization}",
    eprint = "physics/0308063",
    archivePrefix = "arXiv",
    reportNumber = "PHYSTAT-2003-MODT002",
    journal = "eConf",
    volume = "C030908",
    pages = "MODT002",
    year = "2003"
}

@article{holdom1985,
    author = {Holdom, Bob},
    title = {Two U(1)'s and Epsilon Charge Shifts},
    reportNumber = {UTPT-85-30},
    doi = {10.1016/0370-2693(86)91377-8},
    journal = {Phys. Lett. B},
    volume = {166},
    pages = {196--198},
    year = {1986}
}

@article{geant4_2016,
title = {Recent developments in Geant4},
journal = {Nuclear Instruments and Methods in Physics Research Section A: Accelerators, Spectrometers, Detectors and Associated Equipment},
volume = {835},
pages = {186-225},
year = {2016},
issn = {0168-9002},
doi = {https://doi.org/10.1016/j.nima.2016.06.125},
url = {https://www.sciencedirect.com/science/article/pii/S0168900216306957},
author = {J. Allison and others}
}

@ARTICLE{geant4_2006,
  author={Allison, J. and others},
  journal={IEEE Transactions on Nuclear Science}, 
  title={Geant4 developments and applications}, 
  year={2006},
  volume={53},
  number={1},
  pages={270-278}
}

@article{geant4_2003,
title = {Geant4—a simulation toolkit},
journal = {Nuclear Instruments and Methods in Physics Research Section A: Accelerators, Spectrometers, Detectors and Associated Equipment},
volume = {506},
number = {3},
pages = {250-303},
year = {2003},
issn = {0168-9002},
doi = {https://doi.org/10.1016/S0168-9002(03)01368-8},
url = {https://www.sciencedirect.com/science/article/pii/S0168900203013688},
author = {S. Agostinelli and others}
}

@article{alwall2014,
    author = "Alwall, J. and Frederix, R. and Frixione, S. and Hirschi, V. and Maltoni, F. and Mattelaer, O. and Shao, H. -S. and Stelzer, T. and Torrielli, P. and Zaro, M.",
    title = "{The automated computation of tree-level and next-to-leading order differential cross sections, and their matching to parton shower simulations}",
    reportNumber = "CERN-PH-TH-2014-064, CP3-14-18, LPN14-066, MCNET-14-09, ZU-TH-14-14",
    doi = "10.1007/JHEP07(2014)079",
    journal = "JHEP",
    volume = "07",
    pages = "079",
    year = "2014"
}

@article{ambrogi2018,
    author = "Ambrogi, Federico and Arina, Chiara and Backovic, Mihailo and Heisig, Jan and Maltoni, Fabio and Mantani, Luca and Mattelaer, Olivier and Mohlabeng, Gopolang",
    title = "{MadDM v.3.0: a Comprehensive Tool for Dark Matter Studies}",
    reportNumber = "CP3-18-26, MCnet-18-07, MCNET-18-07",
    doi = "10.1016/j.dark.2018.11.009",
    journal = "Phys. Dark Univ.",
    volume = "24",
    pages = "100249",
    year = "2019"
}

@inproceedings{xgboost,
author = {Chen, Tianqi and Guestrin, Carlos},
title = {XGBoost: A Scalable Tree Boosting System},
year = {2016},
isbn = {9781450342322},
publisher = {Association for Computing Machinery},
address = {New York, NY, USA},
url = {https://doi.org/10.1145/2939672.2939785},
doi = {10.1145/2939672.2939785},
booktitle = {Proceedings of the 22nd ACM SIGKDD International Conference on Knowledge Discovery and Data Mining},
pages = {785–794},
numpages = {10},
location = {San Francisco, California, USA},
series = {KDD '16}
}

@article{mu2e2014,
    author = "Bartoszek, L. and others",
    collaboration = "Mu2e Collaboration",
    title = "{Mu2e Technical Design Report}",
    eprint = "1501.05241",
    archivePrefix = "arXiv",
    primaryClass = "physics.ins-det",
    reportNumber = "FERMILAB-TM-2594, FERMILAB-DESIGN-2014-01",
    doi = "10.2172/1172555",
    year = "2014"
}

@article{qu2020,
  title = {Jet tagging via particle clouds},
  author = {Qu, Huilin and Gouskos, Loukas},
  journal = {Phys. Rev. D},
  volume = {101},
  issue = {5},
  pages = {056019},
  numpages = {11},
  year = {2020},
  month = {Mar},
  publisher = {American Physical Society},
  doi = {10.1103/PhysRevD.101.056019},
  url = {https://link.aps.org/doi/10.1103/PhysRevD.101.056019}
}

@article{LESA,
      title={The SLAC Linac to ESA (LESA) Beamline for Dark Sector Searches and Test Beams}, 
      author={Tom Markiewicz and Tor Raubenheimer and Natalia Toro and members of the LESA construction team},
      year={2022},
      eprint={2205.13215},
      archivePrefix={arXiv},
      primaryClass={physics.acc-ph},
      url={https://arxiv.org/abs/2205.13215}, 
}

@phdthesis{horoho2025,
    author = "Horoho, Tyler",
    title = "Searches for Sub-GeV Dark Matter with NOvA and LDMX and Performance Studies of the Cosmic-Ray Veto for Mu2e",
    school = "University of Virginia",
    year = "2025"
}

@article{CMS-HGCal-2017,
    collaboration = "CMS Collaboration",
    title = "{The Phase-2 Upgrade of the CMS Endcap Calorimeter}",
    reportNumber = "CERN-LHCC-2017-023, CMS-TDR-019",
    doi = "10.17181/CERN.IV8M.1JY2",
    year = "2017"
}

@misc{gaiser2026,
      title={Cornering MeV-GeV Axions and Dark Photons with LDMX}, 
      author={Sarah Gaiser and Alessandro Russo and Philip Schuster},
      year={2026},
      eprint={2604.14285},
      archivePrefix={arXiv},
      primaryClass={hep-ph},
      url={https://arxiv.org/abs/2604.14285}, 
}

@article{corrodi2026,
title = {Performance studies of the Mu2e cosmic ray veto detector},
journal = {Nuclear Instruments and Methods in Physics Research Section A: Accelerators, Spectrometers, Detectors and Associated Equipment},
volume = {1083},
pages = {171083},
year = {2026},
issn = {0168-9002},
doi = {https://doi.org/10.1016/j.nima.2025.171083},
url = {https://www.sciencedirect.com/science/article/pii/S016890022500885X},
author = {Simon Corrodi and Mackenzie Devilbiss and E. Craig Dukes and Ralf Ehrlich and R. Craig Group and Tyler Horoho and Yuri Oksuzian and Paul Rubinov and Matthew Solt and Yongyi Wu}
}

\appendix
\appendix

\section{\hcal BDT Features}\label{app:hcal_bdt_feat}

\begin{figure}[ht]
	\centering
	\includegraphics[width=0.48\linewidth] {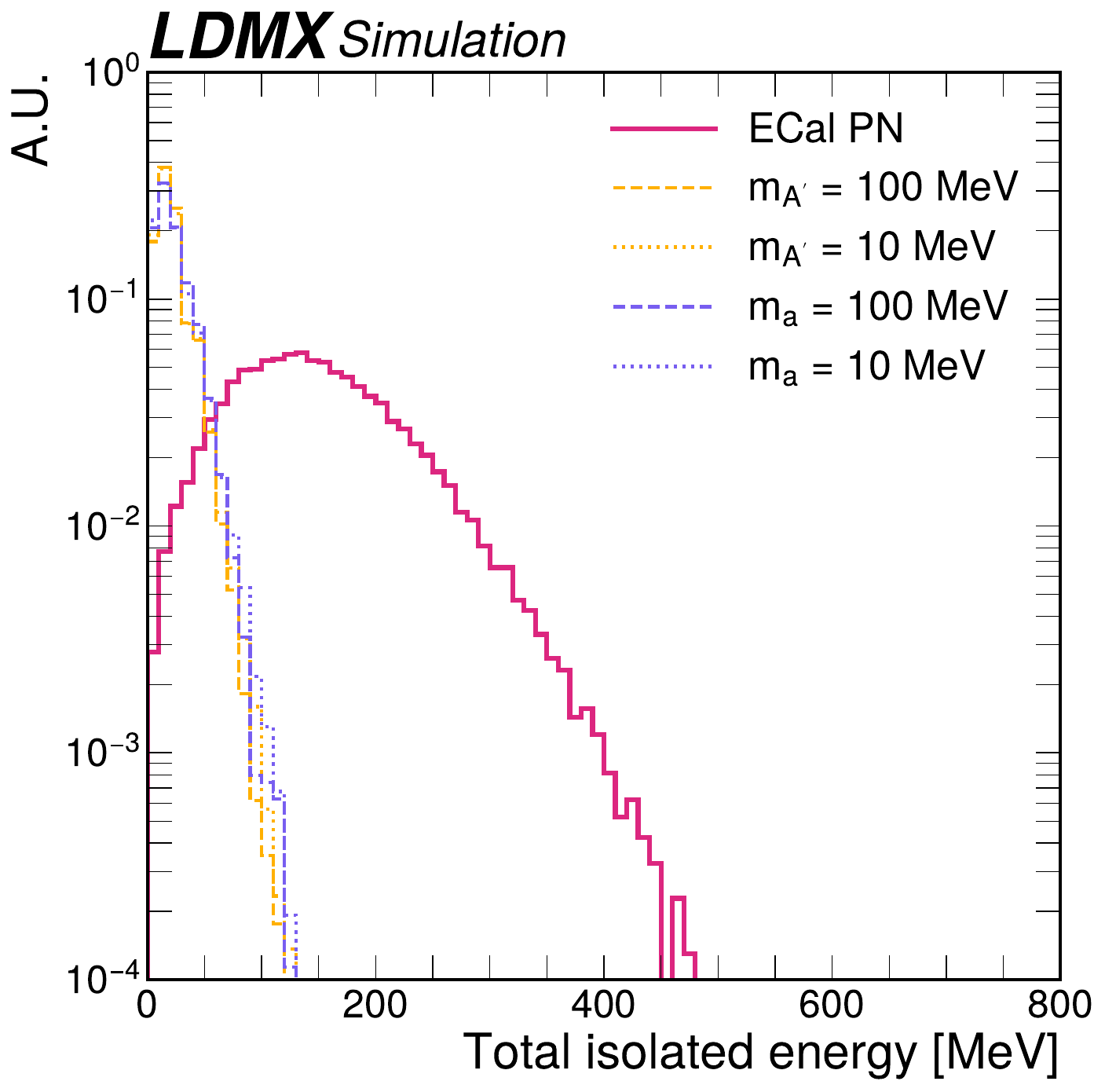}
	\includegraphics[width=0.48\linewidth] {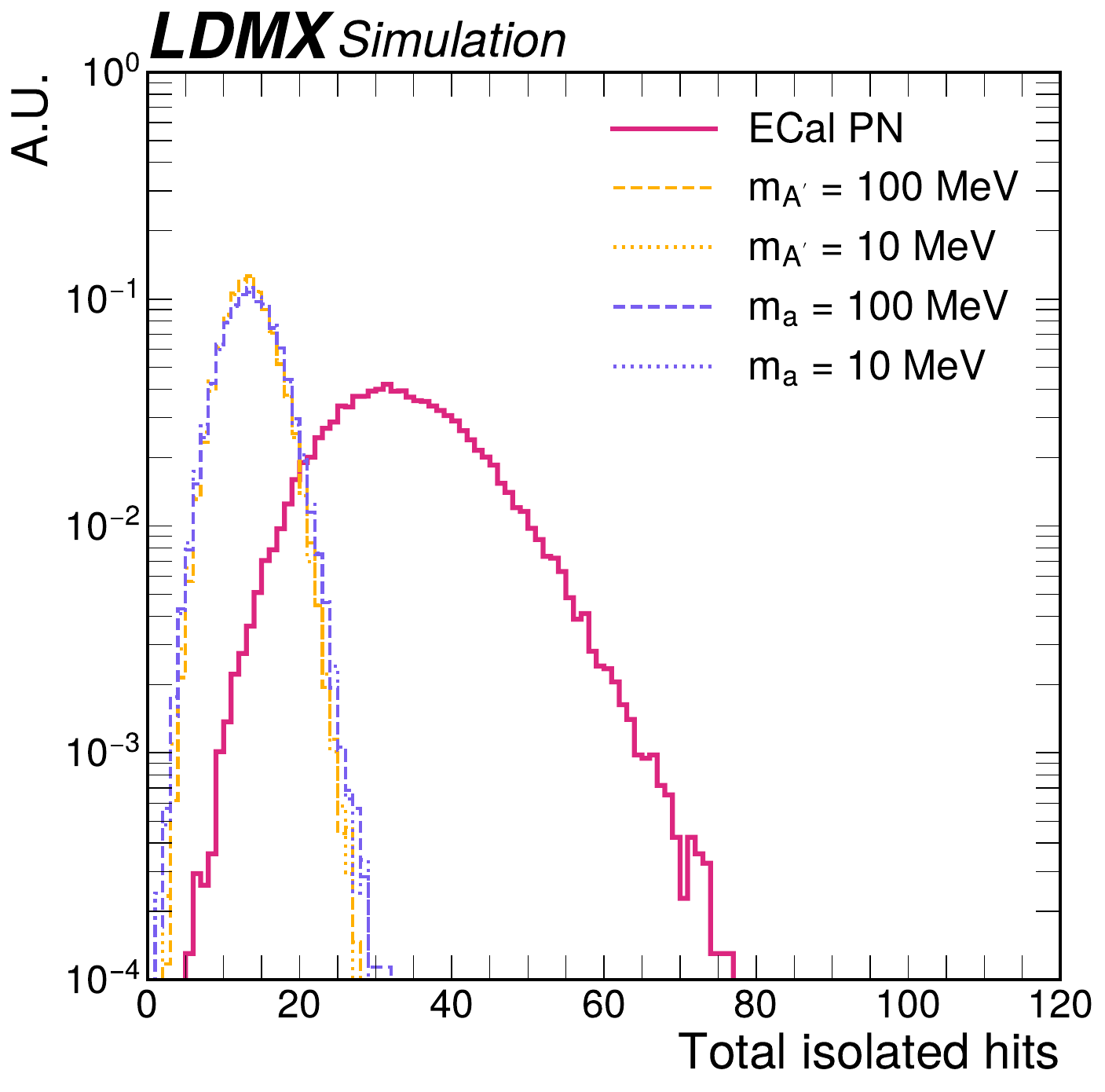} 
    \caption{Features concerning isolated hits in the \hcal: the summed energy in all isolated hits (left) and the number of isolated hits (right).}
\end{figure}

\begin{figure}
    \centering
	\includegraphics[width=0.48\linewidth] {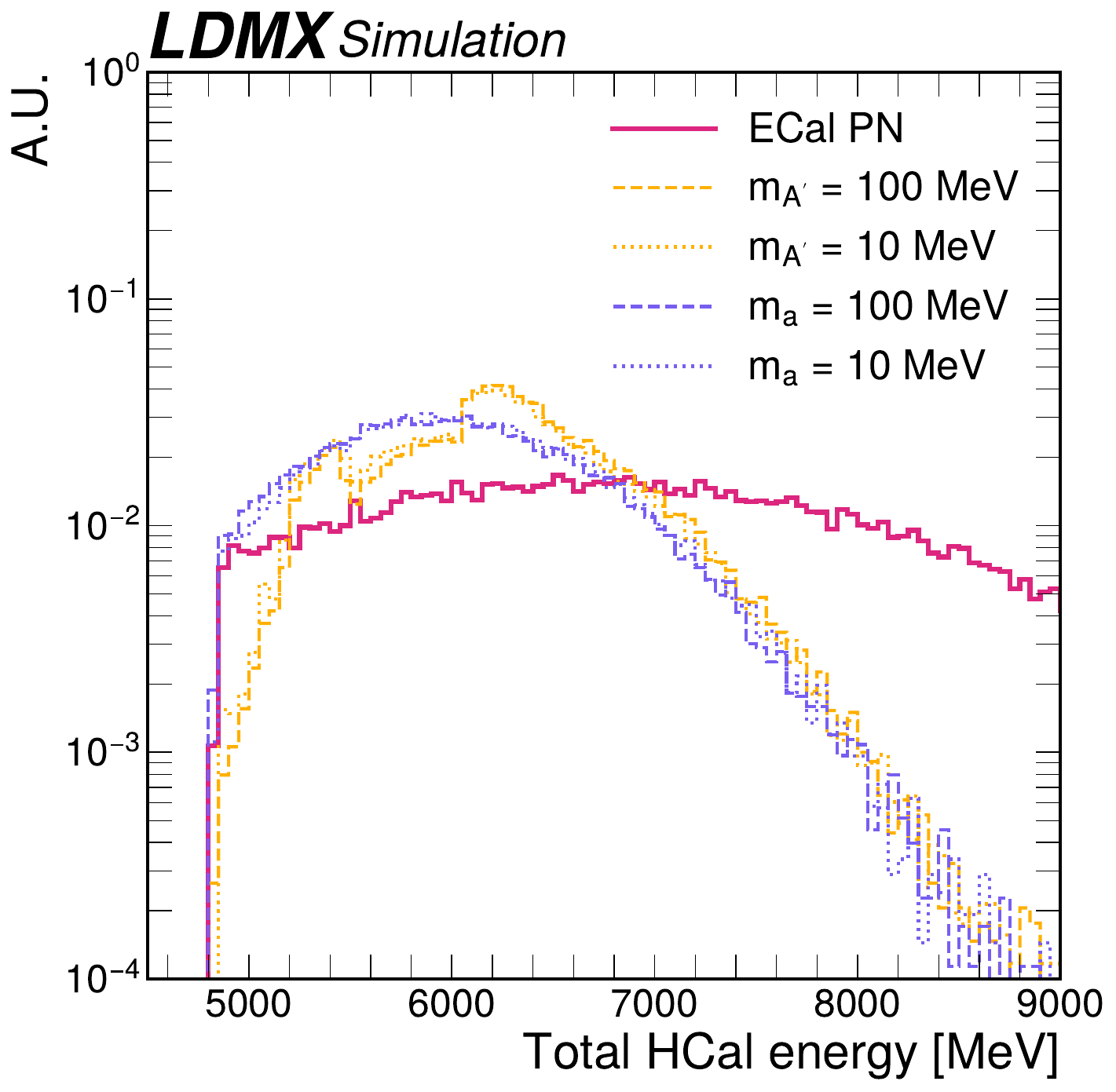}
	\includegraphics[width=0.48\linewidth] {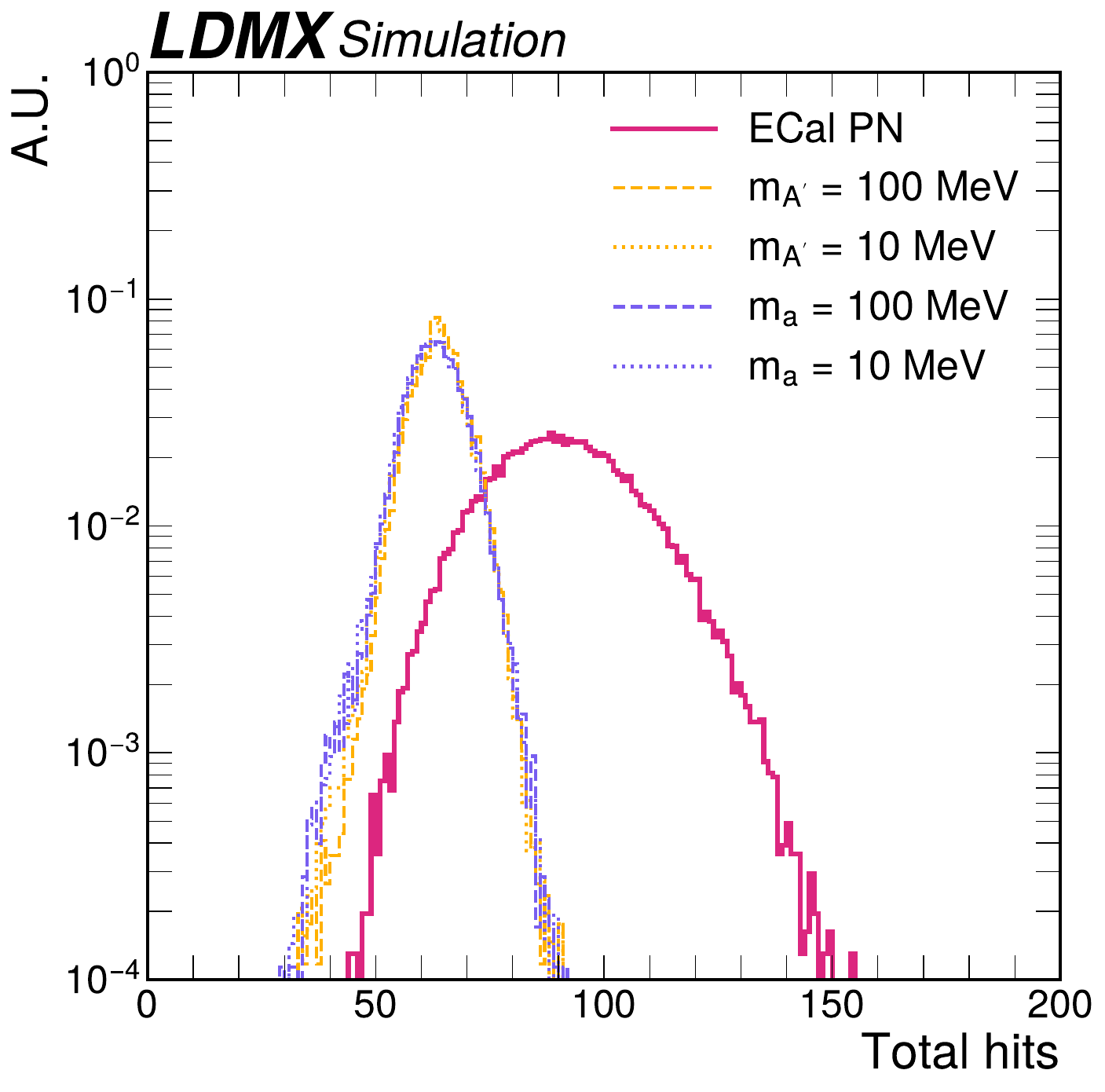}
    \includegraphics[width=0.48\linewidth] {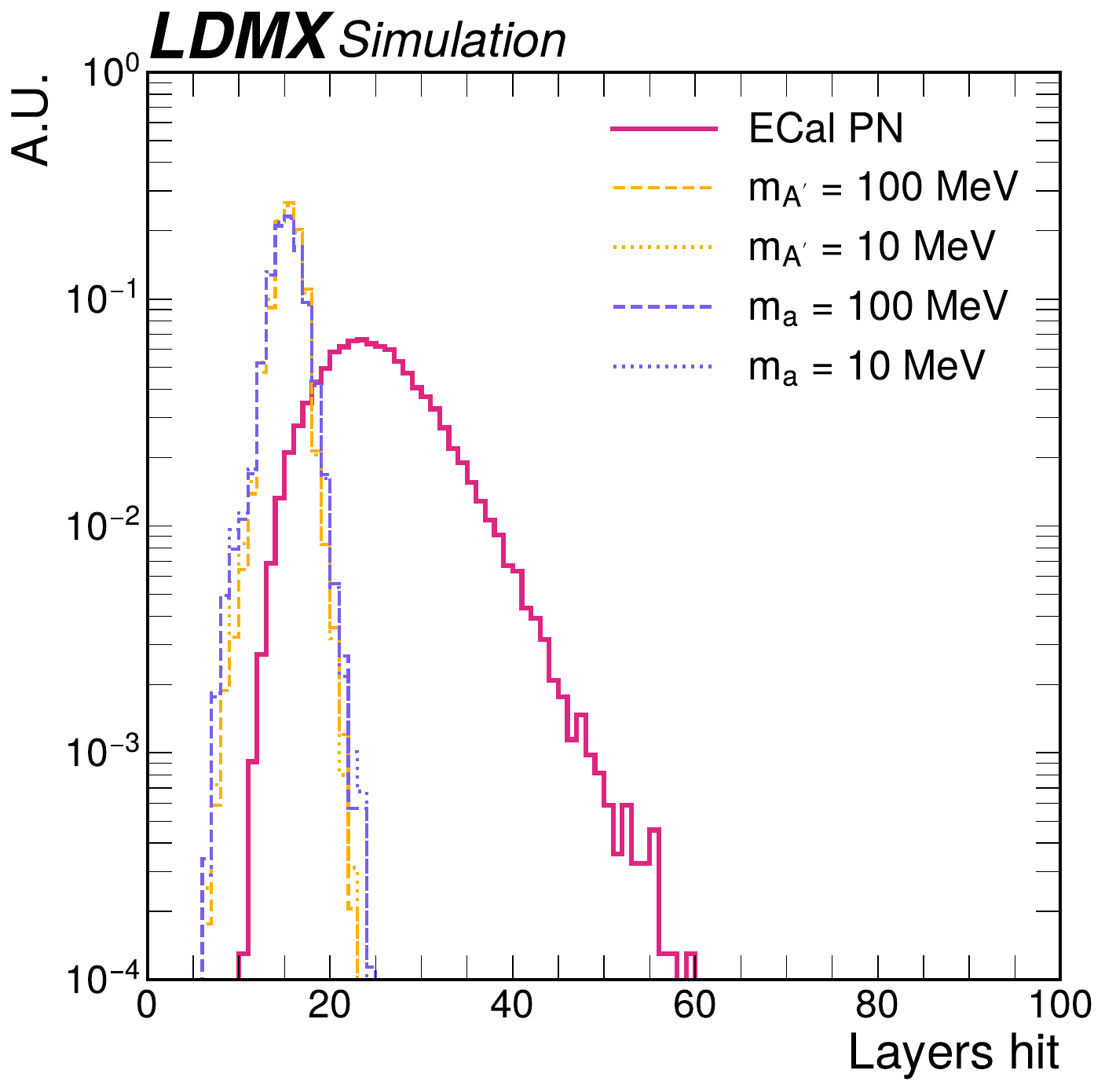}
    \caption{Features concerning all hits in the \hcal: the summed energy in all hits (top left), the total number of hits (top right), and the total number of \hcal layers hit (bottom).}
 \end{figure}

\begin{figure}[ht]
	\centering
	\includegraphics[width=0.48\linewidth]{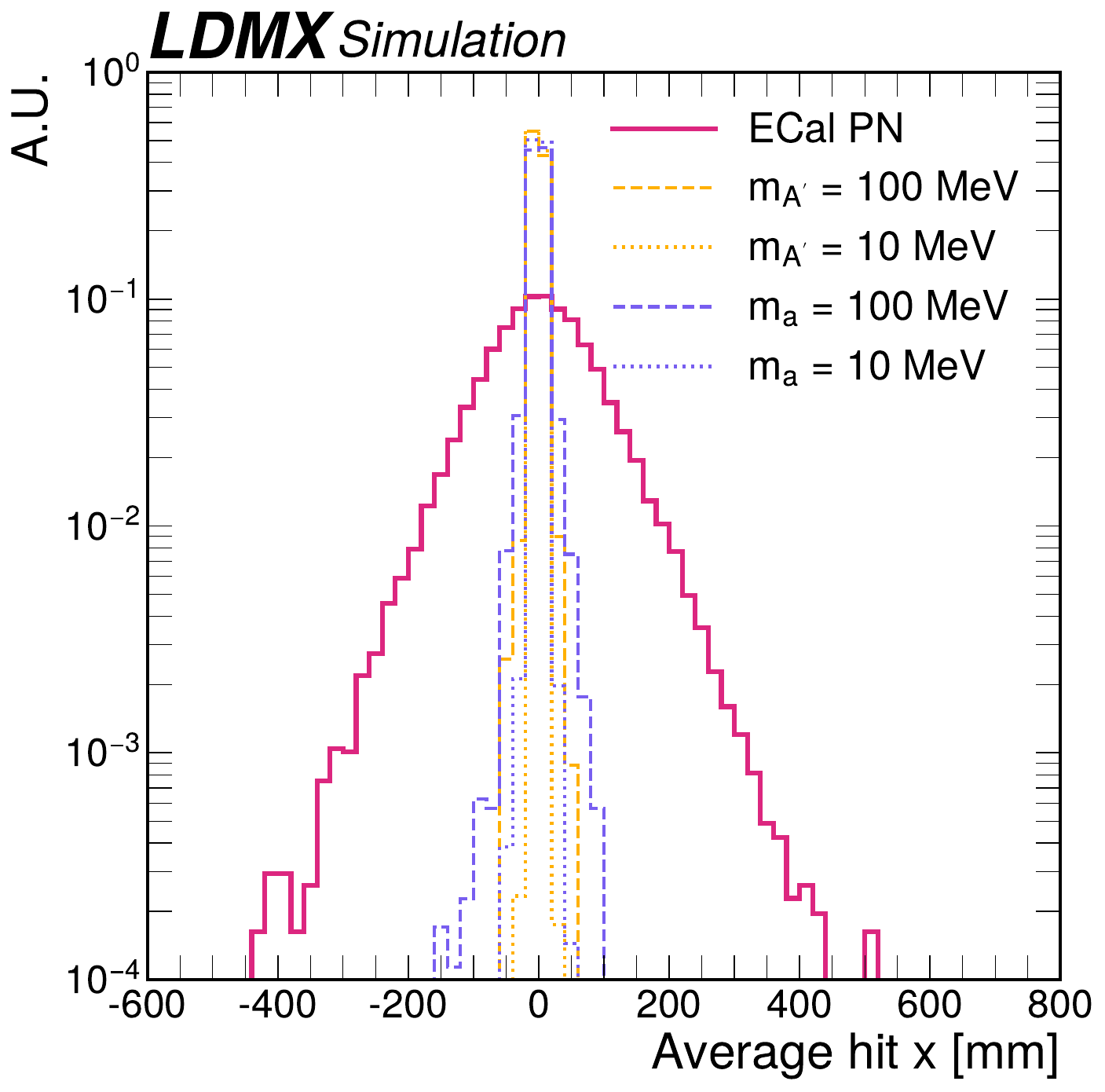} 
	\includegraphics[width=0.48\linewidth] {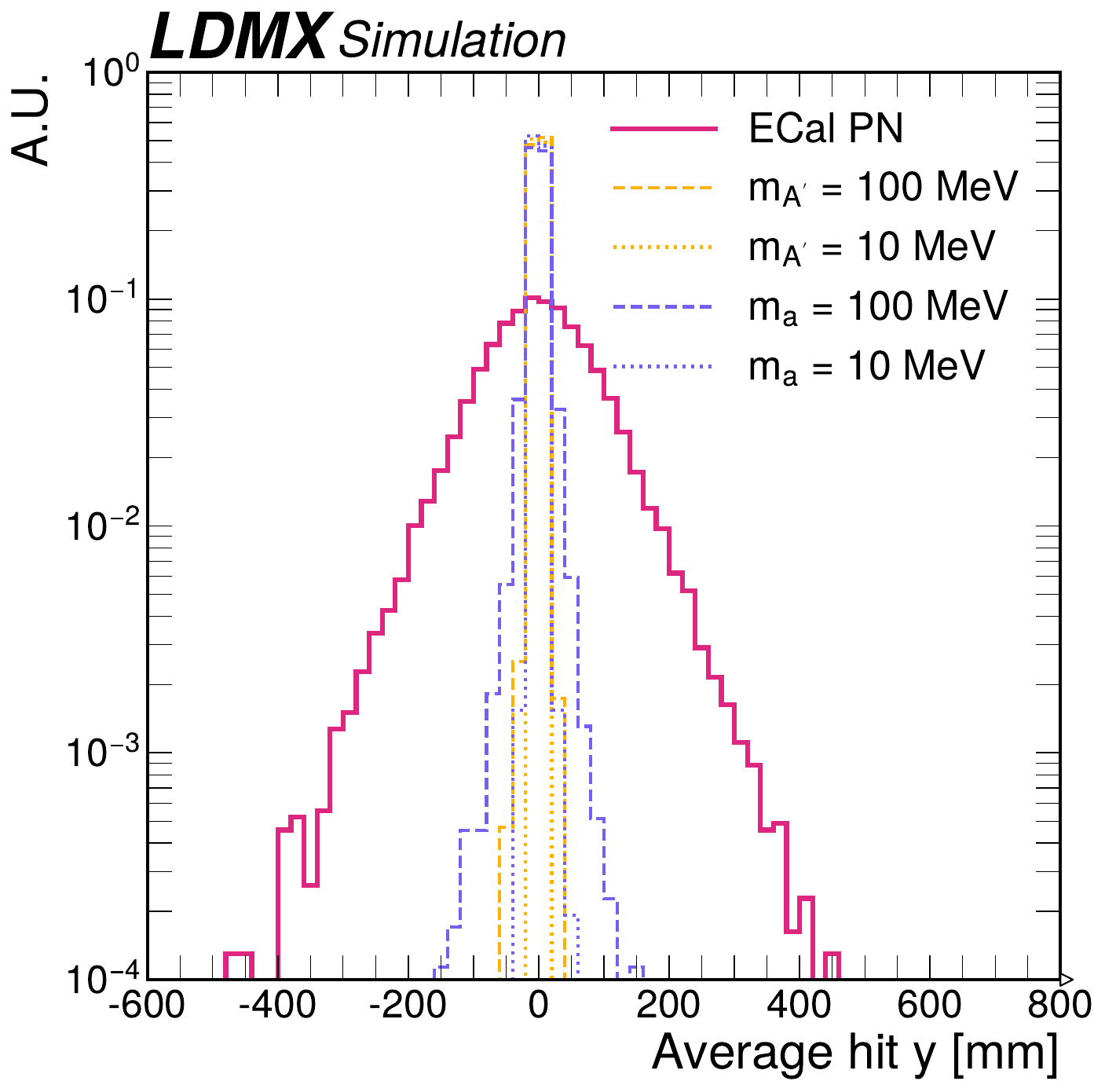}
    \includegraphics[width=0.48\linewidth] {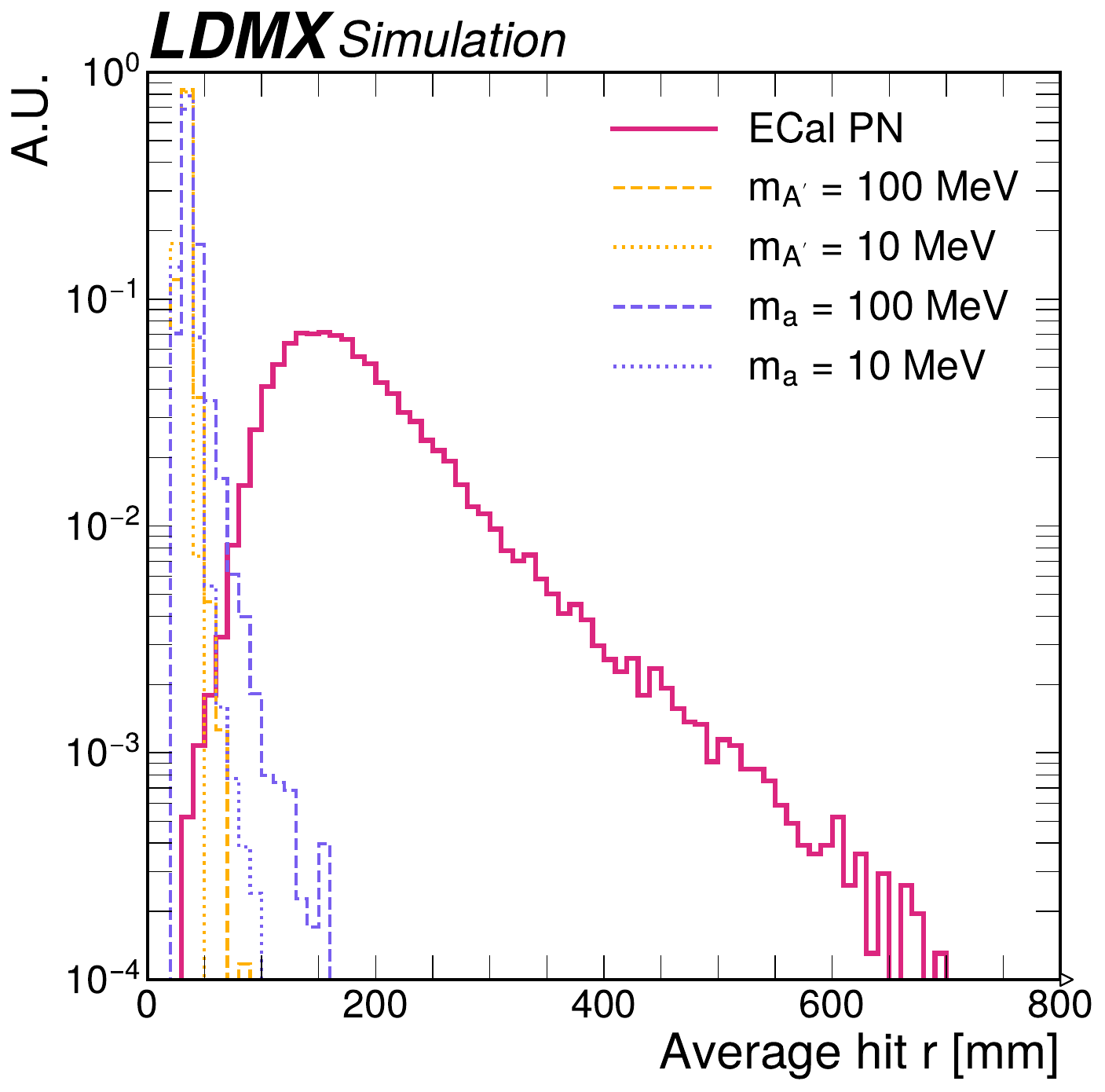}
    \caption{Features for the mean of spatial coordinates: energy-weighted mean hit $x$ (top left), energy-weighted mean hit $y$ (top right), energy-weighted mean hit radius (bottom).}
\end{figure}

\begin{figure}
    \centering
	\includegraphics[width=0.48\linewidth] {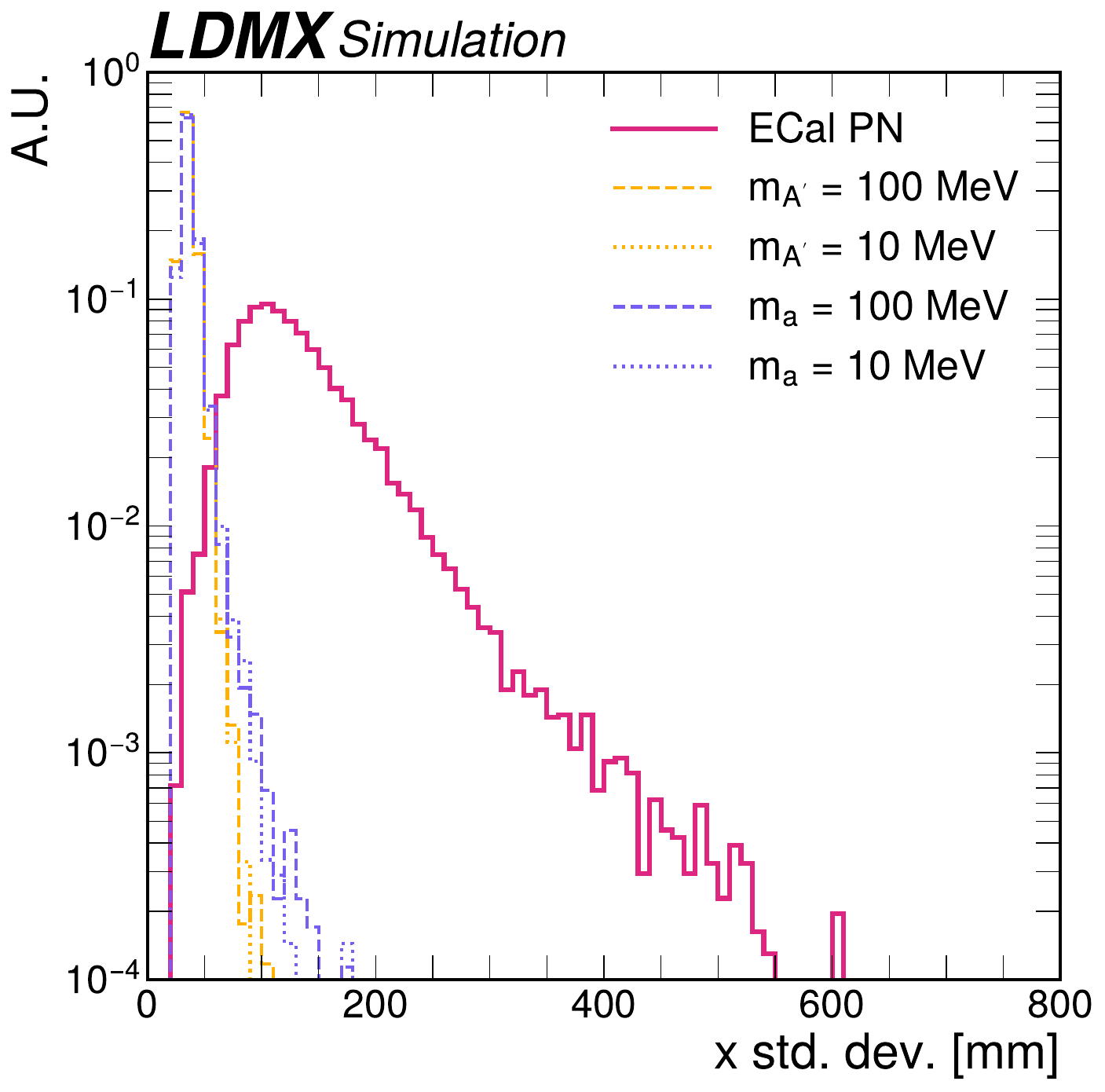} 
	\includegraphics[width=0.48\linewidth] {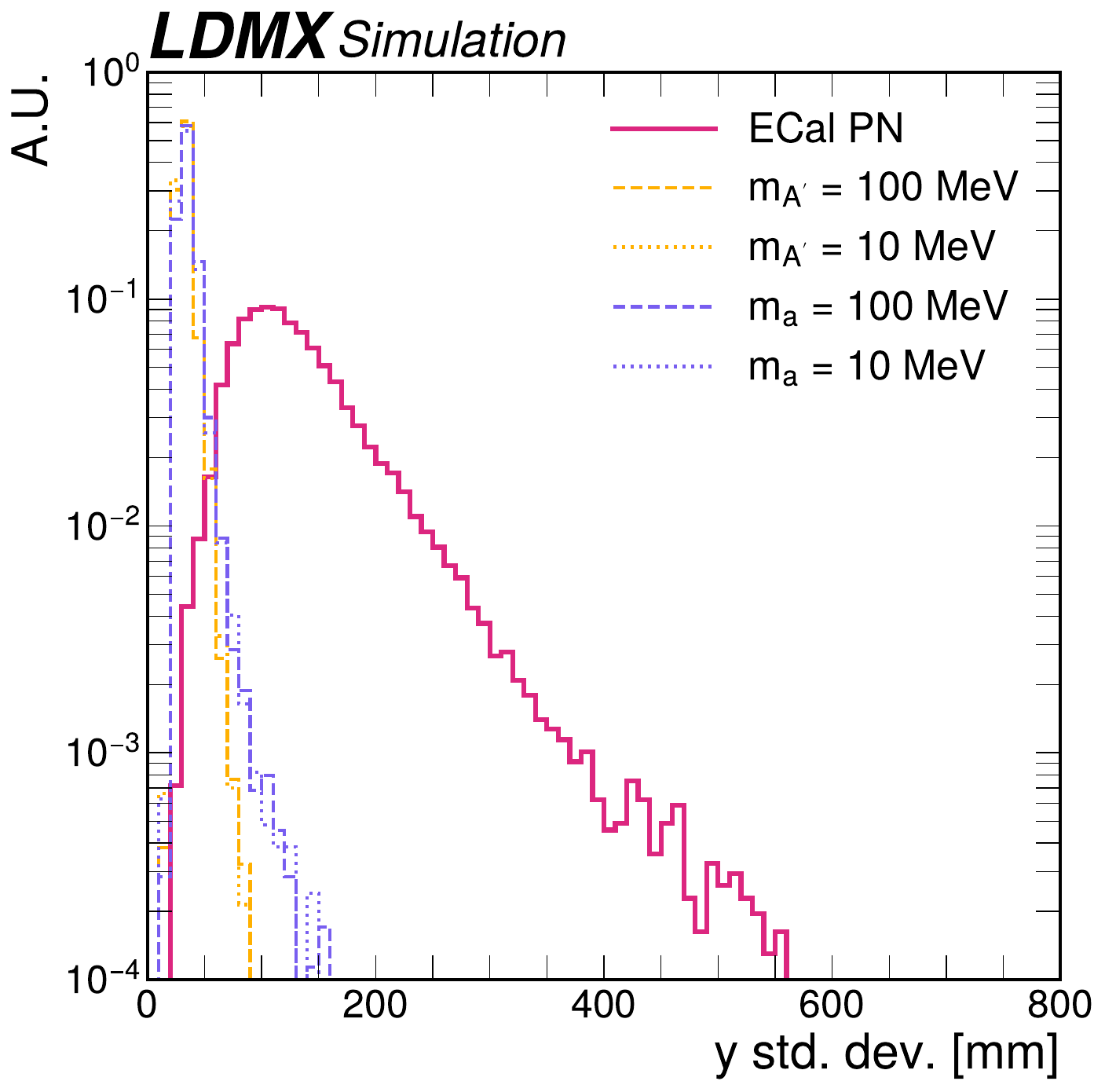}
	\includegraphics[width=0.48\linewidth] {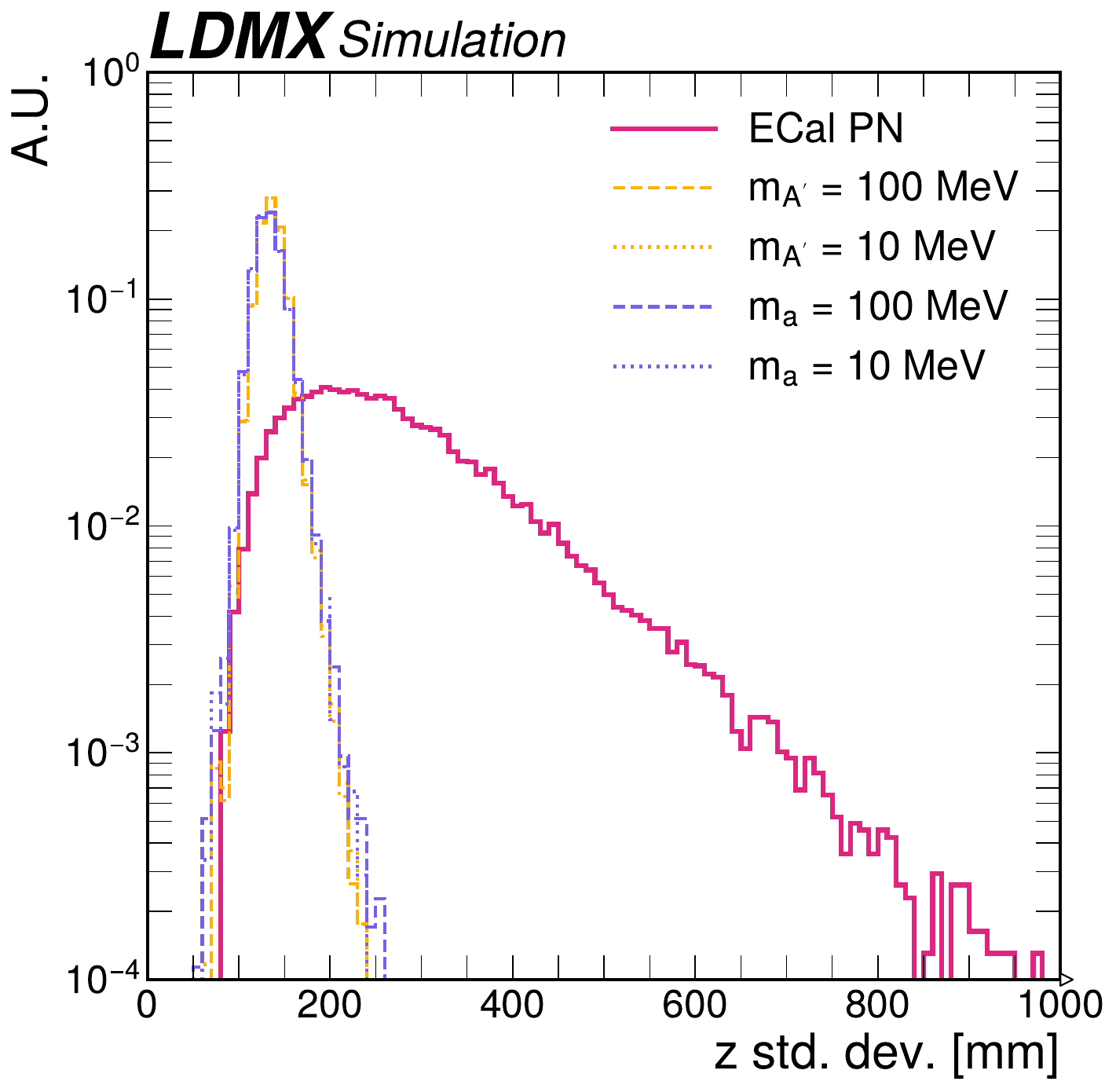} 
    \caption{Features for the standard deviation of spatial coordinates: energy-weighted standard deviation of hits in $x$ (top left), energy-weighted standard deviation of hits in $y$ (top right), energy-weighted standard deviation of hits in $z$ (bottom).}
 \end{figure}

\section{\hcal Signal Efficiency}\label{app:sig_eff_plot}

 \begin{figure}[ht!]
     \centering
     \includegraphics[width=0.6\linewidth]{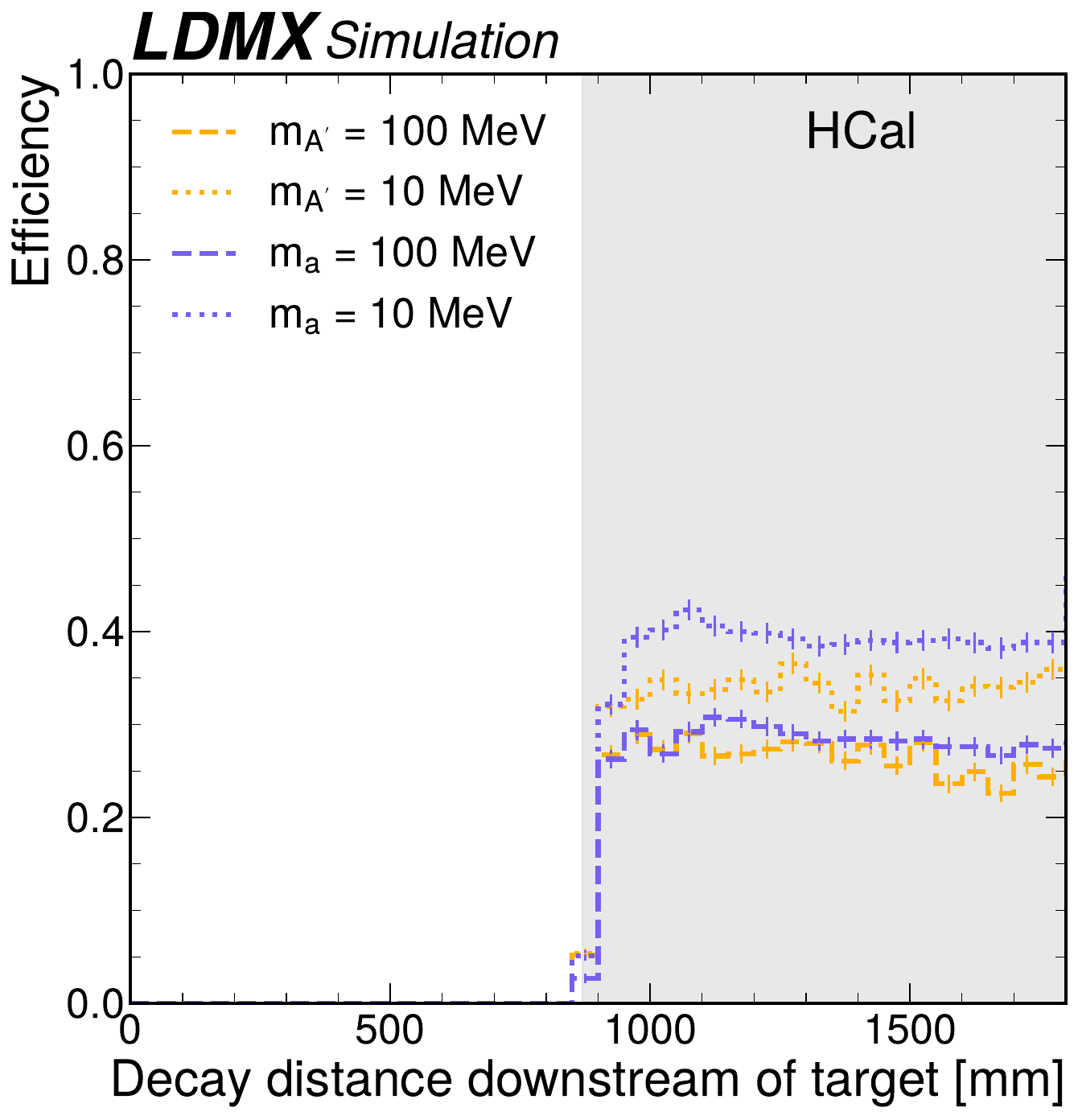}
     \caption{Final signal efficiency as a function of the LLP decay position after all cuts for two $A'$ masses and two ALP masses. The \hcal extends beyond the plotted axis to 5870 mm downstream of the target.}
     \label{fig:signaleff}
\end{figure}

\clearpage
\end{document}